# New insights on the thermal decomposition of lanthanide(III) and actinide(III) oxalates: from neodymium and cerium to plutonium

L. De Almeida,[a] S. Grandjean,*[b] N. Vigier,[c] and F. Patisson[d]

[a] CEA Marcoule, Nuclear Energy Division, DRCP/SCPS/LC2A, Bât. 399, BP 17171, 30207 Bagnols-sur-Cèze, France
Fax: +33 (0)4 66 79 65 67
E-mail: lucie.de-almeida@cea.fr
[b] CEA Marcoule, Nuclear Energy Division, DRCP/DIR, Bât. 400, BP 17171, 30207 Bagnols-sur-Cèze, France
Fax: +33 (0)4 66 79 69 80
E-mail: stephane.grandjean@cea.fr
[c] AREVA NC, BUR/DIRP/RDP, BP 406B, 1 place Jean Millier, 92084 Paris La Défense, France
[d] Institut Jean Lamour, Labex DAMAS, Université de Lorraine, Parc de Saurupt, 54011 Nancy Cedex, France

Lanthanides are often used as surrogates to study the properties of actinide compounds. Their behaviour is considered to be quite similar as they both possess f valence electrons and are close in size and chemical properties. This study examines the potential of using two lanthanides (neodymium and cerium) as surrogates for plutonium during the thermal decomposition of isomorphic oxalate compounds, in the trivalent oxidation state, into oxides. Thus, the thermal decomposition of neodymium, cerium and plutonium (III) oxalates are investigated by several coupled thermal analyses (TG–DTA/DSC–MS/FT-IR) and complementary characterisation techniques (XRD, UV-vis, FT-IR, SEM, carbon analyser) under both oxidizing and inert atmosphere. The thermal decomposition mechanisms determined in this study confirmed some previous results reported in the literature, among diverging propositions, while also elucidating some original mechanisms not previously considered. Calculated thermodynamic and kinetic parameters for the studied systems under both atmospheres are reported and compared with available literature data. Similarities and differences between the thermal behaviour of plutonium(III) and lanthanide(III) oxalates are outlined.

## Introduction

Oxalic conversion, namely precipitation with oxalic acid of cations in solution followed by calcination of the precipitate, is a preferential method to obtain oxides of a wide variety of elements, such as alkaline-earth metals, transition metals, rare-earth metals and actinides.[1],[2] The oxalic conversion route allows not only the quantitative precipitation of cations in solution, but also the control of some physicochemical properties of the resulting oxide via the thermal treatment (morphology, surface area, impurities, etc.).[3]

For instance, the main conversion route of plutonium into oxide is oxalic conversion. Two variants have been industrialised: the first one is based on oxalic precipitation of Pu(IV) and is the most commonly used on an industrial scale (notably in AREVA's spent nuclear fuel recycling plant at La Hague), while the second proceeds via the precipitation of Pu(III) oxalate in reducing conditions. The first route benefits from an industrial background of almost 50 years, whereas the second one allows the formation of an oxide powder suitable for pelletisation which is used in more specific applications (such as $PuO_2$ for sources used in space programs).[4]

This work compares the thermal decomposition path of oxalate compounds containing elements displaying specific redox behaviour during the temperature transitions, especially by evaluating the similarities in the thermal decomposition of trivalent lanthanide and actinide oxalates. Traditionally, the behaviour of lanthanides and actinides is considered to be quite similar as they both possess f valence electrons and are close in size and chemical properties. Lanthanides are thus often used as surrogates to study actinides. Therefore, three elements are examined: neodymium (which commonly exists in the trivalent oxidation state in aqueous solution), cerium (which exists in the trivalent and tetravalent oxidation states) and plutonium (which can display oxidation states from +III to +VII).

The thermal decomposition of either neodymium [5], cerium [6] or plutonium (III) [7] oxalates have been investigated since the 1960's, but the published results tend to differ depending on the reaction intermediates proposed under a given atmosphere, generating a need for clarification. As such, the thermal decomposition mechanisms of the three oxalates were carefully and comparatively investigated in the present study under both inert and oxidizing atmospheres using different techniques. The reaction steps, intermediate products and evolved gases were determined by thermogravimetry (TG), differential thermal analysis (DTA), IR-spectroscopy (FT-IR), mass-spectrometry (MS), X-ray diffraction (XRD) and UV-vis spectroscopy. Scanning electron microscopy (SEM) was used to monitor the progressive changes in morphology of the oxalate precursors throughout their conversion to oxides. More broadly, a methodology for systematic study of solid state transformations of other oxalate compounds is hereby illustrated.

## Results and Discussion

### 1. Thermal decomposition mechanisms of lanthanide(III) oxalates

#### 1.1. Neodymium(III) oxalate

The results of the coupled thermogravimetry-gas analysis studies on the thermal decomposition of neodymium oxalate are presented on Figure 1 under argon atmosphere and on Figure 2 in air. Even though the TG curves under both atmospheres appear to be quite similar, the gases emission profiles as well as the heat flow curves differ perceptibly during the oxalate decomposition phase (starting from about 350°C).

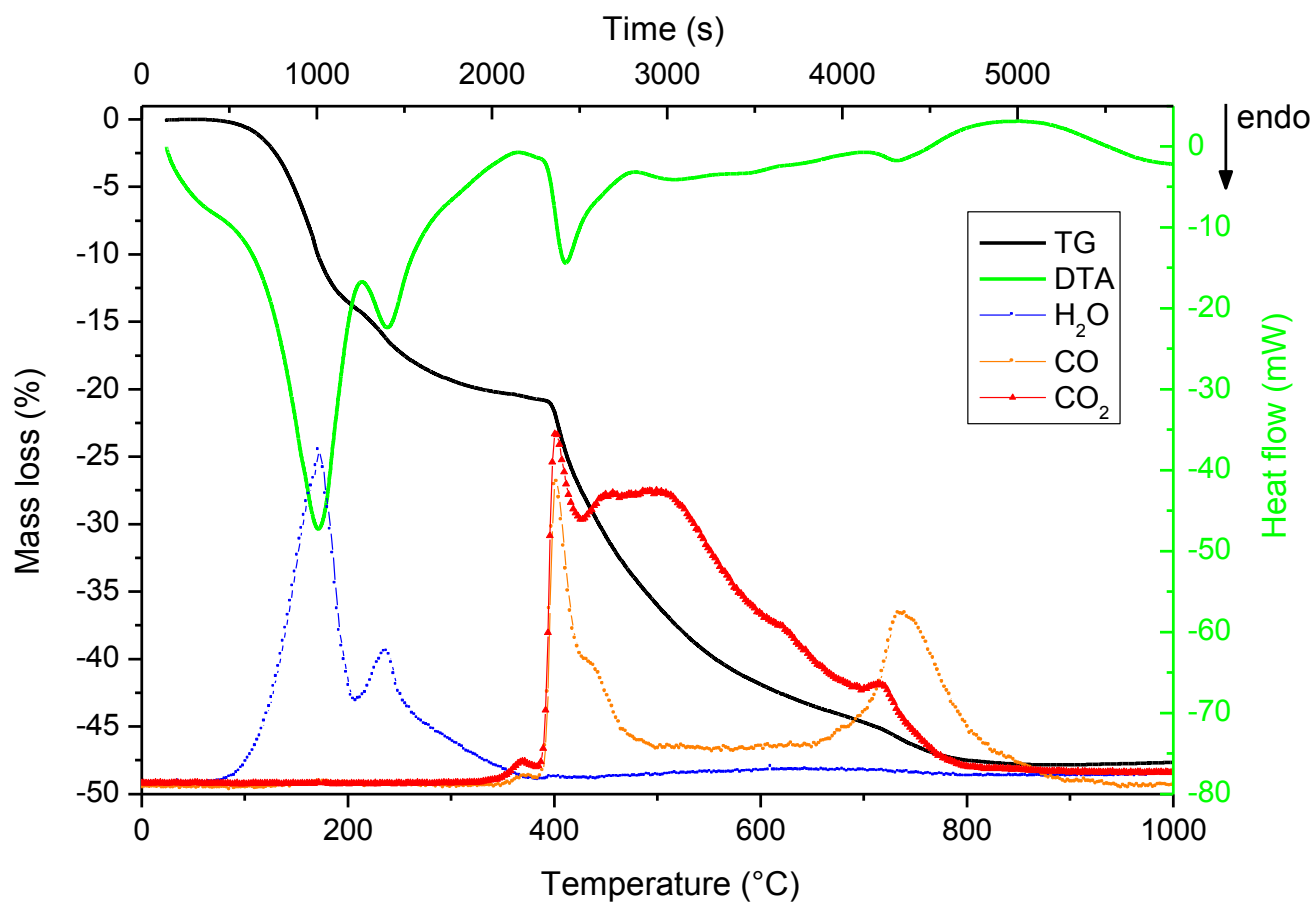

Figure 1 : Thermal decomposition of $Nd_2(C_2O_4)_3 \cdot nH_2O$ under streaming argon (250 mL/min) on heating at 10°C/min – TG, DTA and evolved gases recorded by MS (MS curves : arbitrary units).

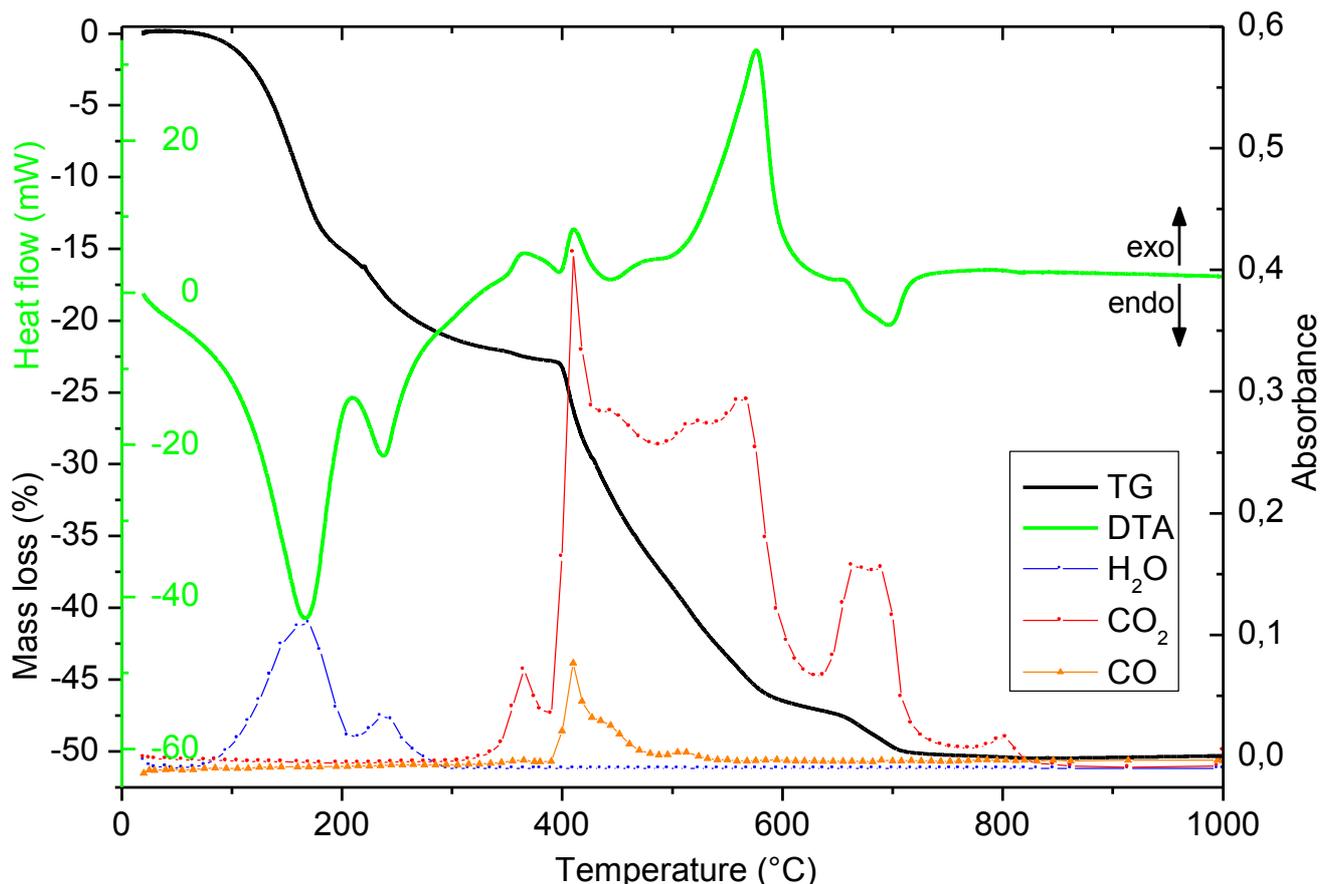

Figure 2 : Thermal decomposition of $Nd_2(C_2O_4)_3 \cdot nH_2O$ in streaming air (250 mL/min) on heating at 10°C/min – TG, DTA and evolved gases recorded by FT-IR.

These plots show that the dehydration occurs in two endothermic steps, regardless of the atmosphere. The first step, centred around 160°C, account for the loss of the first water molecules (in fact six, see below), while the second step, around 240°C, corresponds to the loss of the two remaining water molecules.

Under inert atmosphere, the beginning of the decomposition of the oxalate groups overlaps with the end of the dehydration phase, according to the gases profiles observed with mass spectrometry. Under oxidizing atmosphere, the dehydration and decomposition phases are distinct. The oxalate decomposition leads to the release of CO and $CO_2$, occurring in three exothermic steps and one endothermic step in air, as opposed to three endothermic steps under argon.

Under oxidizing atmosphere, CO is mostly oxidized by $O_2$. On the contrary, CO partially disproportionates under inert atmosphere (around 400-500°C) to form $CO_2$ and elemental carbon (Boudouard reaction), causing the powder to darken. Unexpectedly, the last step of the thermal decomposition under argon (around 650-850°C) involves the release of CO as well as $CO_2$: this comes from the Boudouard equilibrium which is displaced above 650°C toward the consumption of residual carbon and $CO_2$ to form CO (the thermodynamic inversion temperature in the Ellingham diagram is 700°C). The oxide obtained after calcination up to 1000°C under argon atmosphere is black, indicating the presence of residual elemental carbon.

Tables 1 and 2 compare experimental and theoretical mass losses, corresponding to the TG analyses of neodymium oxalate under argon and in air. It appeared that the starting oxalate dehydrated in the thermobalance prior to the experiment, because of the vacuum used to empty the device before filling with the required atmosphere, so the dehydration steps only account for eight water molecules.

Table 1. Comparison between the experimental and theoretical mass losses for the calcination of neodymium oxalate under argon at 10°C/min

| | Proposed reaction step | Temperature range (°C) | Theoretical mass loss (%) | Experimental mass loss (%) |
|---|---|---|---|---|
| 1 | $Nd_2(C_2O_4)_3 \cdot 8H_2O \rightarrow Nd_2(C_2O_4)_3 \cdot 2H_2O$ | 90 – 200 | 15.51 | 15.46 |
| 2 | $Nd_2(C_2O_4)_3 \cdot 2H_2O \rightarrow Nd_2(C_2O_4)_3$ | 200 – 350 | 5.17 | 5.19 |
| 3 | $Nd_2(C_2O_4)_3 \rightarrow Nd_2(CO_3)_3$ | 350 – 480 | 12.06 | 11.93 |
| 4 | $Nd_2(CO_3)_3 \rightarrow Nd_2O_2CO_3$ | 480 – 650 | 12.63 | 11.85 |
| 5 | $Nd_2O_2CO_3 \rightarrow Nd_2O_3$ | 650 – 850 | 6.32 | 3.79 |

The difference between the experimental and theoretical mass losses for the last step of the thermal decomposition under argon is due to the presence of residual carbon in the oxide, as explained in **§** *1.3*.

Table 2. Comparison between the experimental and theoretical mass losses for the calcination of neodymium oxalate in air at 5°C/min

| | Proposed reaction step | Temperature range (°C) | Theoretical mass loss (%) | Experimental mass loss (%) |
|---|---|---|---|---|
| 1 | $Nd_2(C_2O_4)_3 \cdot 8H_2O \rightarrow Nd_2(C_2O_4)_3 \cdot 2H_2O$ | 90 – 190 | 15.51 | 15.52 |
| 2 | $Nd_2(C_2O_4)_3 \cdot 2H_2O \rightarrow Nd_2(C_2O_4)_3$ | 200 – 325 | 5.17 | 5.18 |
| 3 | $Nd_2(C_2O_4)_3 \rightarrow Nd_2(C_2O_4)_{5/2}(CO_3)_{1/2}$ | 325 – 400 | 2.01 | 2.05 |
| 4 | $Nd_2(C_2O_4)_{5/2}(CO_3)_{1/2} \rightarrow Nd_2(CO_3)_3$ | 400 – 425 | 10.05 | 10.08 |
| 5 | $Nd_2(CO_3)_3 \rightarrow Nd_2O_2CO_3$ | 425 – 600 | 12.63 | 12.61 |
| 6 | $Nd_2O_2CO_3 \rightarrow Nd_2O_3$ | 600 – 800 | 6.32 | 6.26 |

Complementary analyses were carried out in order to clarify the decomposition mechanisms, by characterising the reaction intermediates observed on the TG curves. Table 3 presents the attribution of the infrared absorption bands of interest.

Table 3. Attribution of the infrared absorption bands of interest [8]

| Functional group | ν (cm⁻¹) | Intensity | Vibration |
|---|---|---|---|
| $H_2O$ | 2600-3660 | m | $\nu_s$ OH |
| | 1610 | m | $\delta$ HOH |
| $C_2O_4^{2-}$ | 1600 | m | $\nu_{as}$ CO |
| | 1460 | s | $\nu_s$ CO |
| | 1350 | m | $\nu_s$ CO + $\nu$ CC |
| | 1300 | m | $\nu_s$ CO + $\delta$ OCO |
| | 940 | s | $\nu_s$ CO + $\delta$ OCO |
| | 905 | s | $\nu$ CC |
| | 790 | m | $\delta$ OCO + $\nu$ MO |
| $CO_3^{2-}$ | 1510 | m | $\nu_{as}$ CO |
| | 1360 | m | $\nu_{as}$ CO |
| | 1100 | s | $\nu_s$ CO |
| | 900 | m | $\pi$ CO$_3$ |
| | 820 | s | $\delta$ OCO |
| $CO_2$ | 2350 | m | $\nu_{as}$ CO |

Figures 3 to 5 show the results of the FT-IR and HT-XRD analyses of the intermediates compounds under argon and in air.

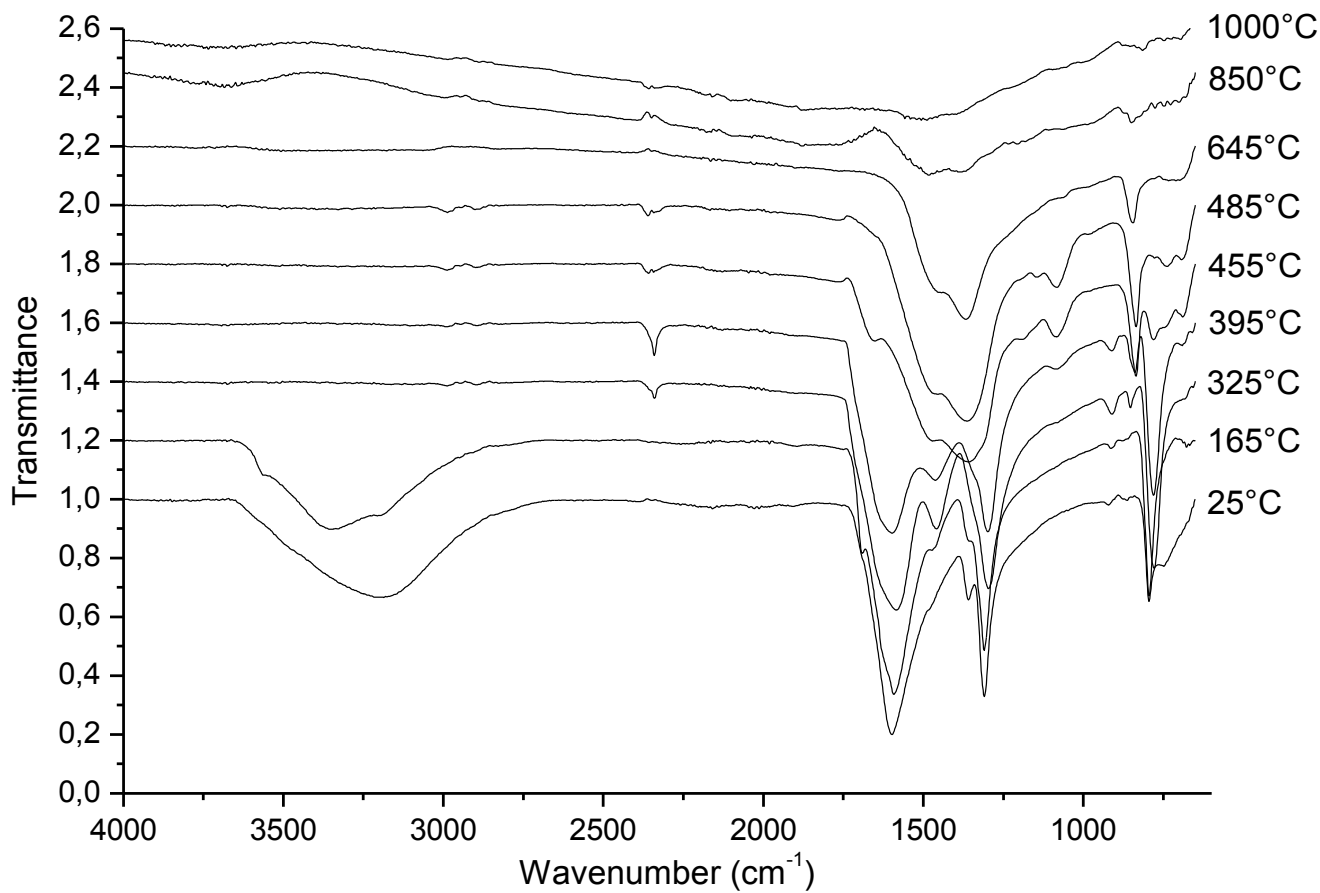

Figure 3 : FT-IR spectra of $Nd_2(C_2O_4)_3 \cdot nH_2O$ thermal decomposition intermediates under argon.

The water vibration signal (broad band between 2700 and 3600 cm$^{-1}$) has disappeared at 325°C, while the oxalate group absorption bands broaden (1590, 1460, 1300, 910, 850 et 785 cm$^{-1}$). The characteristic signal of sorbed $CO_2$ (2340 cm$^{-1}$) shows in the spectra of the intermediates calcined at 325 and 395°C. The appearance of weak carbonate bands at 1080 and 840 cm$^{-1}$ demonstrates that the oxalate groups start to decompose at 395°C. The oxalate bands have totally disappeared at 485°C, then the carbonate bands decrease between 645 and 850°C and disappear completely in the residual oxide spectrum.

High temperature X-ray diffraction under inert atmosphere (see supporting information) reveals a structural change around 80°C between neodymium oxalate hydrate (JCPDS 00-020-0764) and an intermediate structure not described in the literature, seemingly the dihydrate, which would be consistent with the TGA. The compound becomes amorphous with the beginning of the second dehydration step around 200°C. Cubic neodymium oxide (JCPDS 00-021-0579) forms at 600°C and the rise in temperature leads to a phase transition toward hexagonal neodymium oxide (JCPDS 00-041-1089), which stays incomplete even after calcination up to 1000°C (this might be due to slow kinetics).

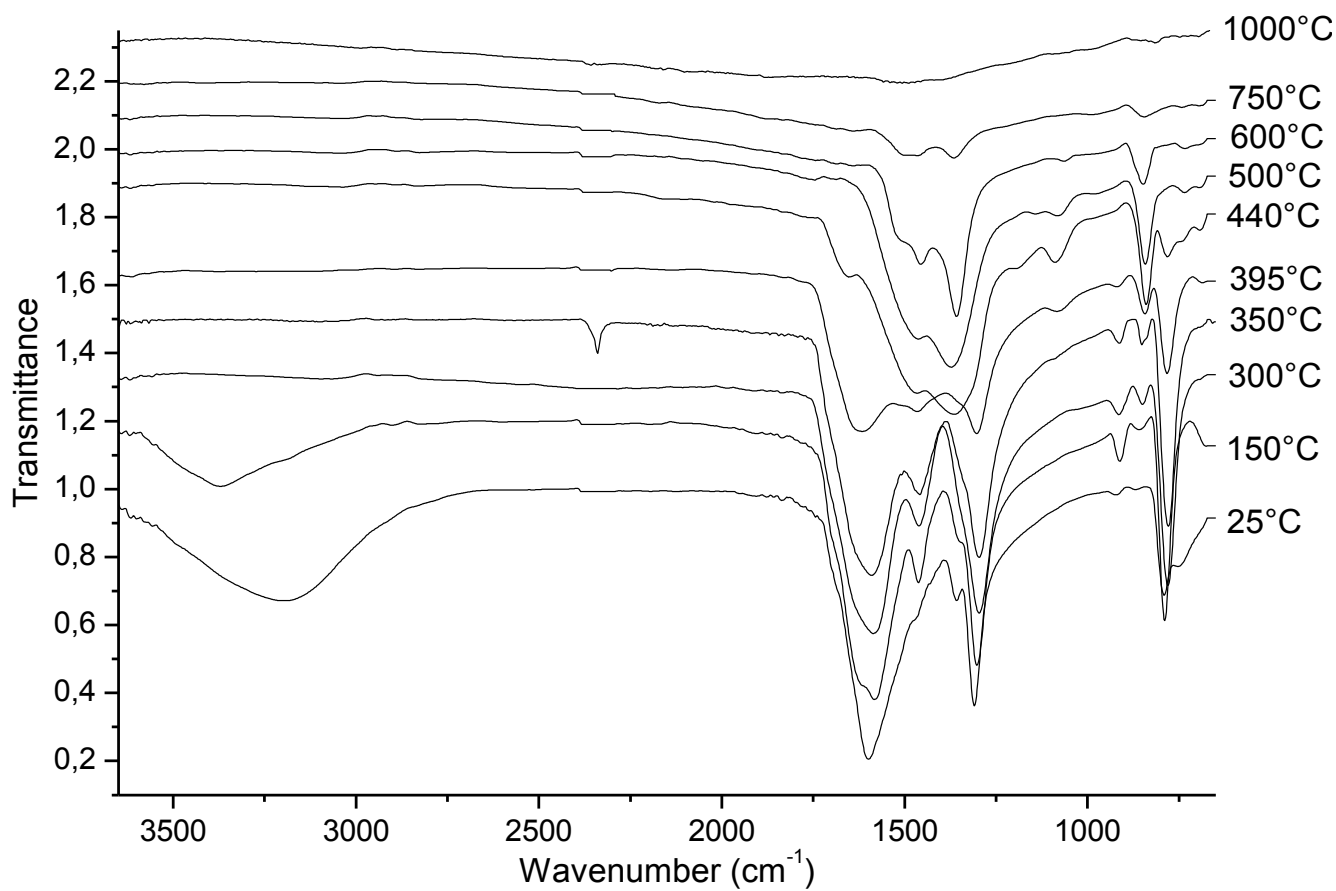

Figure 4 : FT-IR spectra of $Nd_2(C_2O_4)_3 \cdot nH_2O$ thermal decomposition intermediates in air.

Infrared spectroscopy of the intermediates in air (Figure 4) confirms that the compound is totally dehydrated by 300°C. The sorbed $CO_2$ signal appears at 350°C, and the IR spectrum of the intermediate calcined at 395°C shows absorption bands of both oxalate and carbonate groups, supporting the hypothesis of an oxalato-carbonate intermediate. Similar spectra were only reported previously for the thermal decomposition of plutonium(IV) in air, and attributed to the formation of several oxalato-carbonate intermediates.[9] Above 400°C the oxalate bands disappear completely, while the carbonate bands intensify and then decrease till above 750°C.

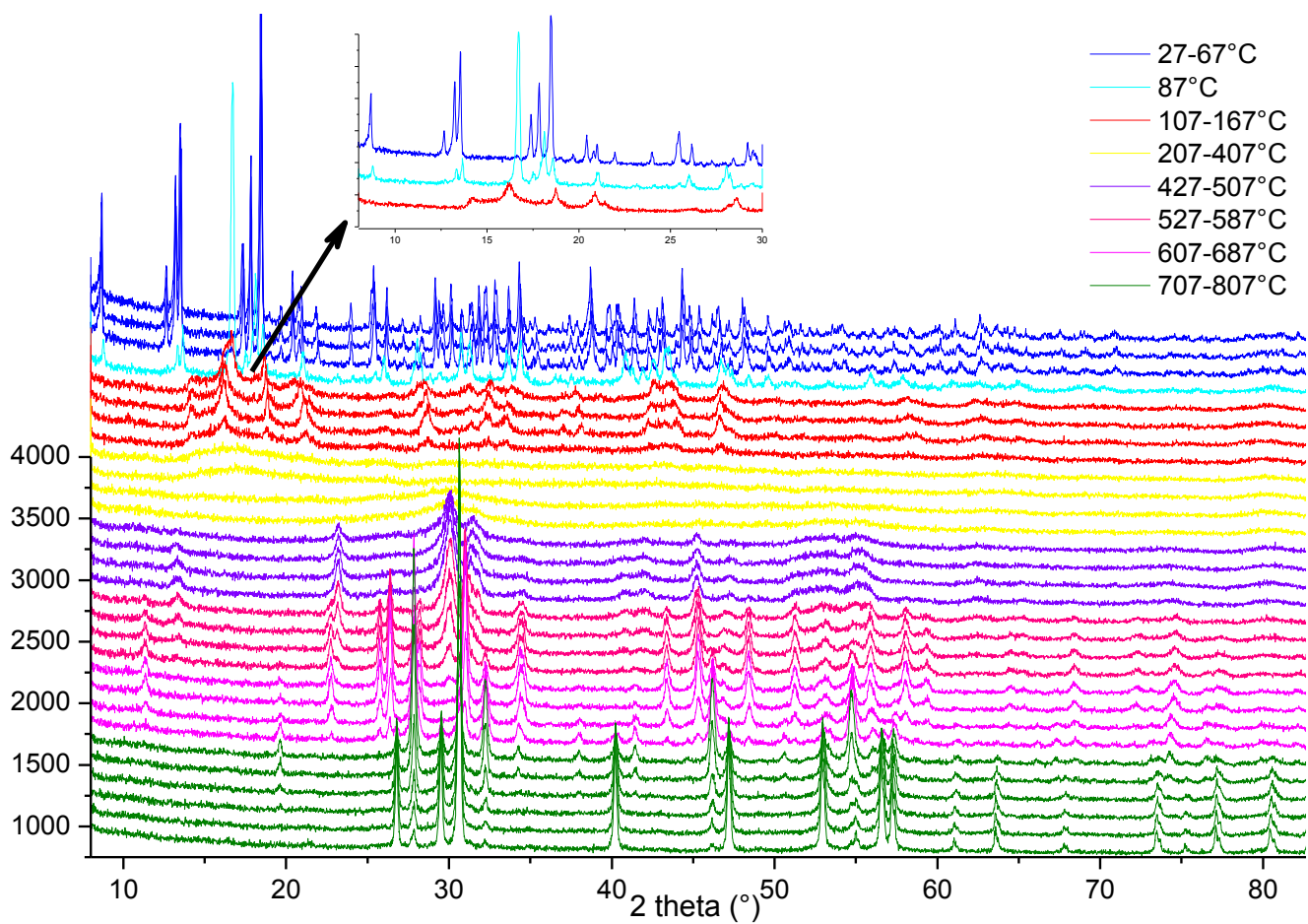

Figure 5 : HT-XRD diffractograms of $Nd_2(C_2O_4)_3 \cdot nH_2O$ and its decomposition in air.

HT-XRD of neodymium oxalate in air (Figure 5) underlines the formation of several phases during the calcination. The crystallographic structure of the initial oxalate changes after 90°C toward that of the same intermediate structure as under air, seemingly the dihydrate, then the solid becomes amorphous from 200°C to 400°C. Around 425°C, a tetragonal intermediate $Nd_2O_2CO_3$ (JCPDS 00-025-0567) appears and persists up to 600°C, followed by a hexagonal $Nd_2O_2CO_3$ phase (JCPDS 00-037-0806) from 525°C to 690°C. Cubic $Nd_2O_3$ forms around 600°C and shifts progressively toward hexagonal $Nd_2O_3$, which is the only phase detected after calcination at 1000°C.

The combination of these varied analysis techniques allows to propose the following mechanisms for the thermal decomposition of neodymium oxalate under argon (i) and air (ii).

(i) Under argon:

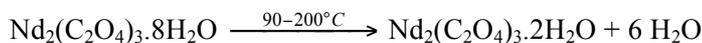

$Nd_2(C_2O_4)_3 \cdot 8H_2O \xrightarrow{90-200°C} Nd_2(C_2O_4)_3 \cdot 2H_2O + 6\ H_2O$

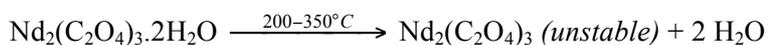

$Nd_2(C_2O_4)_3 \cdot 2H_2O \xrightarrow{200-350°C} Nd_2(C_2O_4)_3\ (unstable) + 2\ H_2O$

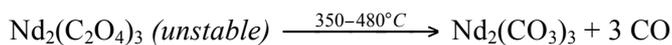

$Nd_2(C_2O_4)_3\ (unstable) \xrightarrow{350-480°C} Nd_2(CO_3)_3 + 3\ CO$

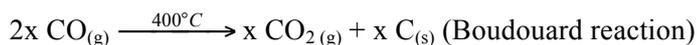

$2x\ CO_{(g)} \xrightarrow{400°C} x\ CO_{2\ (g)} + x\ C_{(s)}$ (Boudouard reaction)

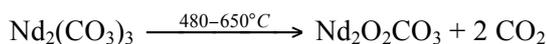

$Nd_2(CO_3)_3 \xrightarrow{480-650°C} Nd_2O_2CO_3 + 2\ CO_2$

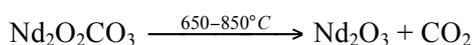

$Nd_2O_2CO_3 \xrightarrow{650-850°C} Nd_2O_3 + CO_2$

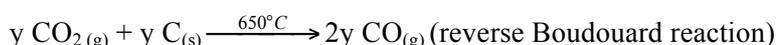

$y\ CO_{2\ (g)} + y\ C_{(s)} \xrightarrow{650°C} 2y\ CO_{(g)}$ (reverse Boudouard reaction)

In addition to the oxide, (x-y) mole of elemental carbon remains (cf. § *1.3*).

(ii) In air:

$$Nd_2(C_2O_4)_3 \cdot 8H_2O \xrightarrow{90-190°C} Nd_2(C_2O_4)_3 \cdot 2H_2O + 6\ H_2O$$

$$Nd_2(C_2O_4)_3 \cdot 2H_2O \xrightarrow{200-325°C} Nd_2(C_2O_4)_3 + 2\ H_2O$$

$$Nd_2(C_2O_4)_3 \xrightarrow{325-400°C} Nd_2(C_2O_4)_{5/2}(CO_3)_{1/2} + 0.5\ CO$$

$$\tfrac{1}{2}\ CO + \tfrac{1}{4}\ O_2 \xrightarrow{325-400°C} \tfrac{1}{2}\ CO_2$$

$$Nd_2(C_2O_4)_{5/2}(CO_3)_{1/2} \xrightarrow{400-425°C} Nd_2(CO_3)_3 + 2.5\ CO$$

$$2x\ CO + x\ O_2 \xrightarrow{400-450°C} 2x\ CO_2$$

$$Nd_2(CO_3)_3 \xrightarrow{425-600°C} Nd_2O_2CO_3 + CO_2$$

$$Nd_2O_2CO_3 \xrightarrow{600-700°C} Nd_2O_3 + CO_2$$

The carbon monoxide evolved during the decomposition of oxalate groups is oxidized by air and is no longer detected above 450°C.

According to Hinode *et al.* [10], the decomposition process of neodymium carbonate takes place as follow:

$Nd_2(CO_3)_3 \rightarrow (NdO)_2CO_3$–*tetragonal* $\rightarrow Nd_2O_3$–*cubic*
$\qquad\qquad\qquad\downarrow\qquad\qquad\qquad\downarrow$
$\qquad\quad(NdO)_2CO_3$–*hexagonal* $\rightarrow Nd_2O_3$–*hexagonal*.

This pattern is confirmed by HT-XRD for the thermal decomposition of neodymium oxalate under oxidizing atmosphere. As the intermediate products appear amorphous under nitrogen, we can't conclude that the thermal decomposition of neodymium oxalate under inert atmosphere follows the same process.

### 1.2. Cerium oxalate decomposition

The results of the coupled TG – DTA – MS/FT-IR studies of the thermal decomposition of cerium oxalate are presented on Figure 6 under argon and on Figure 7 in air. The influence of the calcination atmosphere (oxidizing or inert) appears clearly: the decomposition is complete at 850°C under argon as opposed to 450°C in air.

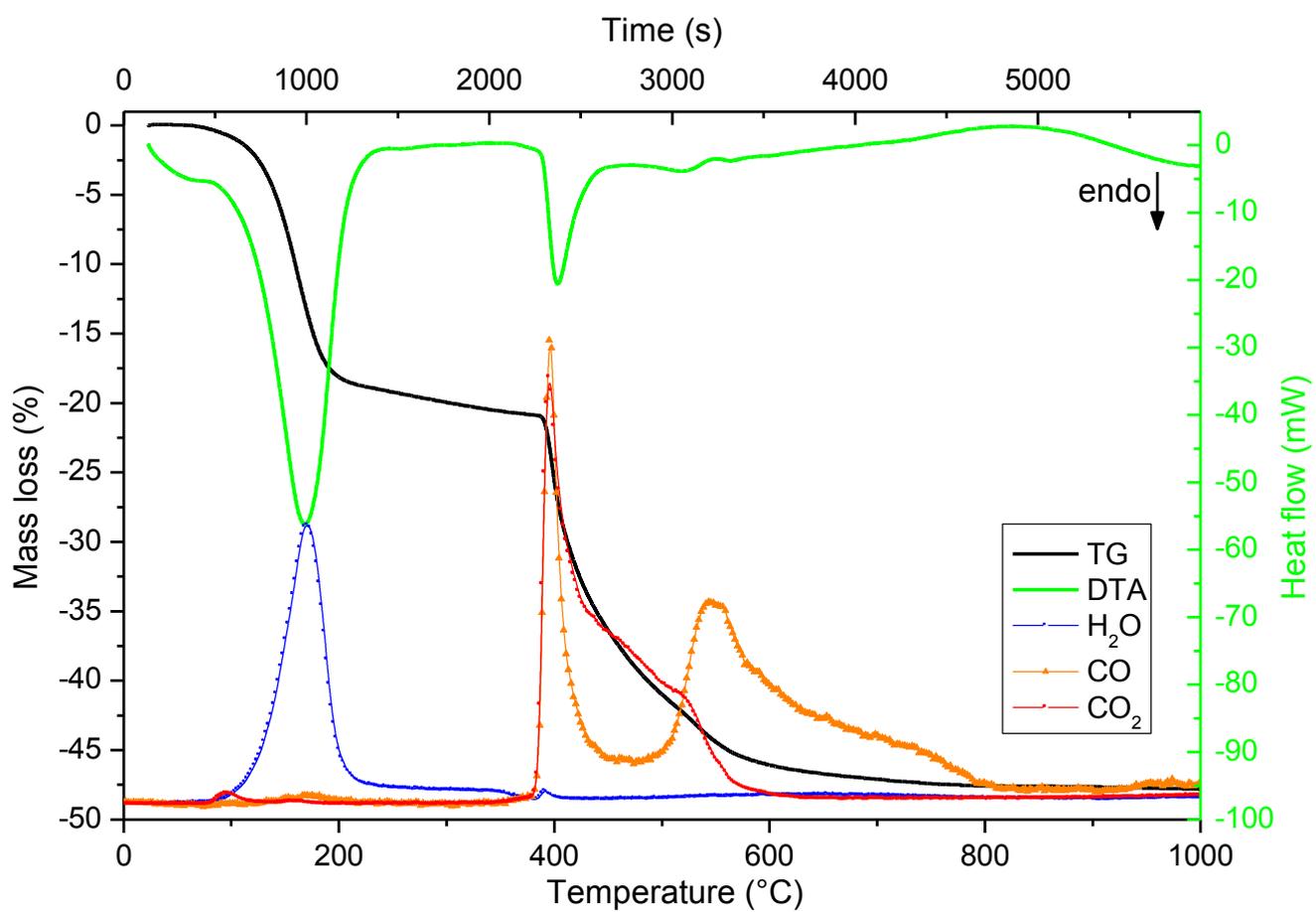

Figure 6 : Thermal decomposition of $Ce_2(C_2O_4)_3 \cdot nH_2O$ under streaming argon (250 mL/min) on heating at 10°C/min – TG, DTA and evolved gases recorded by MS (arbitrary units).

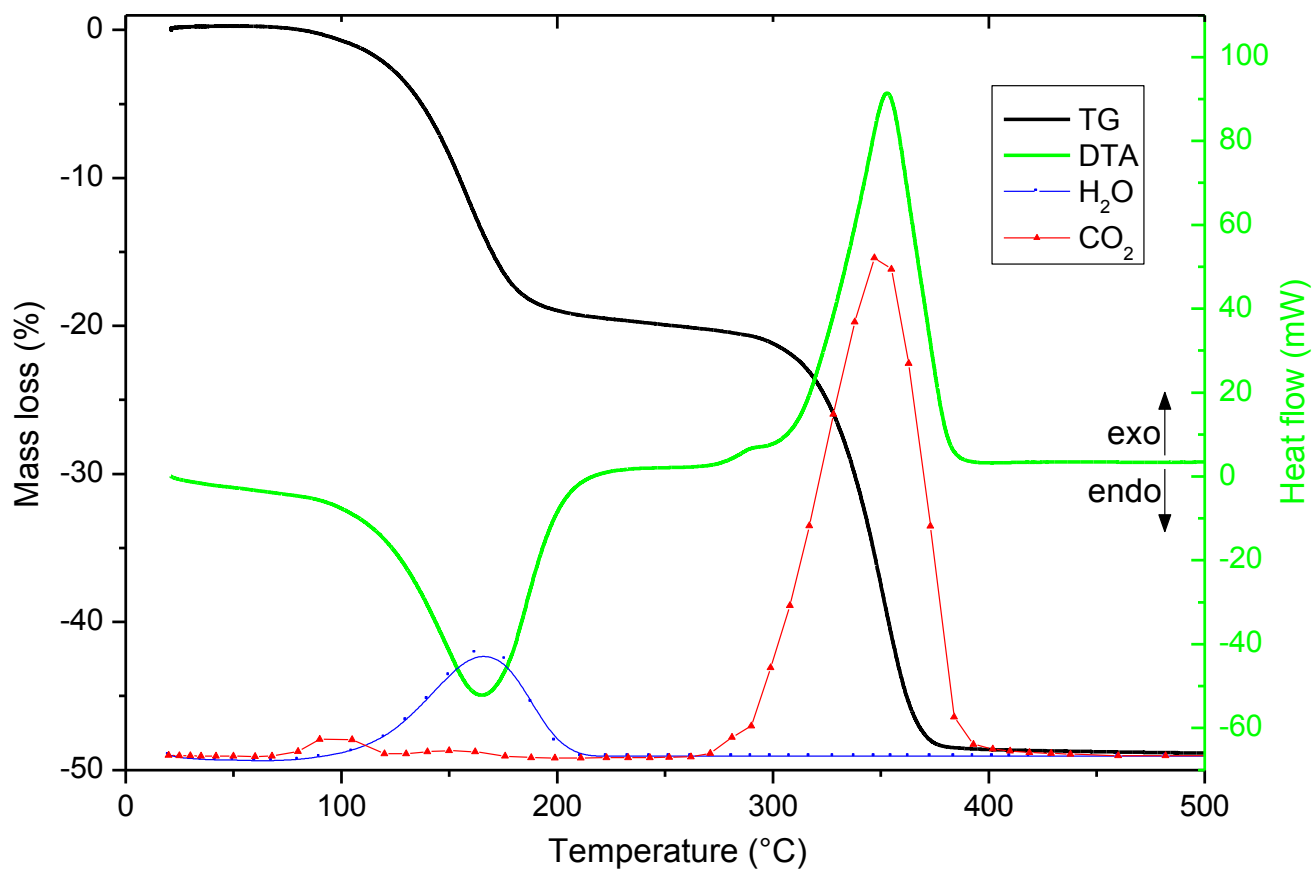

Figure 7 : Thermal decomposition of $Ce_2(C_2O_4)_3 \cdot nOH_2O$ under streaming air (250 mL/min) on heating at 10°C/min – TG, DTA and evolved gases recorded by FT-IR.

These plots show that regardless of the atmosphere, the dehydration of cerium oxalate hydrate occurs in one single step centred around 170°C. The anhydrous oxalate is stable over a substantial range of temperature, then the oxalate groups start to decompose differently depending on the atmosphere. Under inert atmosphere, this decomposition takes place in two steps between 350°C and 800°C, and leads to the release of $CO_2$ and CO. On the other hand, the decomposition is complete in one step between 270°C and 450°C under oxidizing atmosphere, and only evolved $CO_2$ is detected.

While CO is completely oxidized in air, it is detected under argon atmosphere. As for the thermal decomposition of neodymium oxalate, CO partially disproportionates during the thermal decomposition of cerium oxalate under inert atmosphere (around 400-475°C) to form $CO_2$ and elemental carbon (Boudouard reaction), causing the powder to darken. Above 500°C, the Boudouard equilibrium is reversed, residual carbon and $CO_2$ being consumed to form CO. Mechanisms and kinetics of the reverse Boudouard reaction catalysed over metal carbonates have been studied by Nagase *et al.* [11], and Osaki & Mori [12] showed that catalysed reverse Boudouard reaction could occur at 500°C. The oxide powder obtained after calcination up to 1000°C under argon atmosphere is black, indicating the presence of residual elemental carbon.

Tables 4 and 5 compare experimental and theoretical mass losses, corresponding to the TG analyses of cerium oxalate under argon and in air. It appeared once again that the oxalate precursor self-dehydrated prior to the measurement so the starting oxalate only contains eight water molecules.

Table 4. Comparison between the experimental and theoretical mass losses for the calcination of cerium oxalate under argon at 2°C/min

| | Proposed reaction step | Temperature range (°C) | Theoretical mass loss (%) | Experimental mass loss (%) |
|---|---|---|---|---|
| 1 | $Ce_2(C_2O_4)_3 \cdot 8H_2O \rightarrow Ce_2(C_2O_4)_3$ | 90 – 300 | 20.86 | 20.90 |
| 2 | $Ce_2(C_2O_4)_3 \rightarrow Ce_2O_2CO_3$ | 350 – 500 | 24.92 | 23.47 |
| 3 | $Ce_2O_2CO_3 \rightarrow 2\ CeO_2$ | 500 – 800 | 4.06 | 4.14 |

The difference between the experimental and theoretical mass losses for the second step of the thermal decomposition under argon is due to the formation of elemental carbon as soon as the decomposition of the oxalate groups starts, as explained in § *1.3*.

Table 5. Comparison between the experimental and theoretical mass losses for the calcination of cerium oxalate in air at 10°C/min

| | Proposed reaction step | Temperature range (°C) | Theoretical mass loss (%) | Experimental mass loss (%) |
|---|---|---|---|---|
| 1 | $Ce_2(C_2O_4)_3 \cdot 8H_2O \rightarrow Ce_2(C_2O_4)_3$ | 90 – 220 | 20.86 | 20.74 |
| 2 | $Ce_2(C_2O_4)_3 \rightarrow 2\ CeO_2$ | 270 – 450 | 28.98 | 28.97 |

Complementary analyses were carried out on specific reactions intermediates in order to substantiate the decomposition mechanisms.

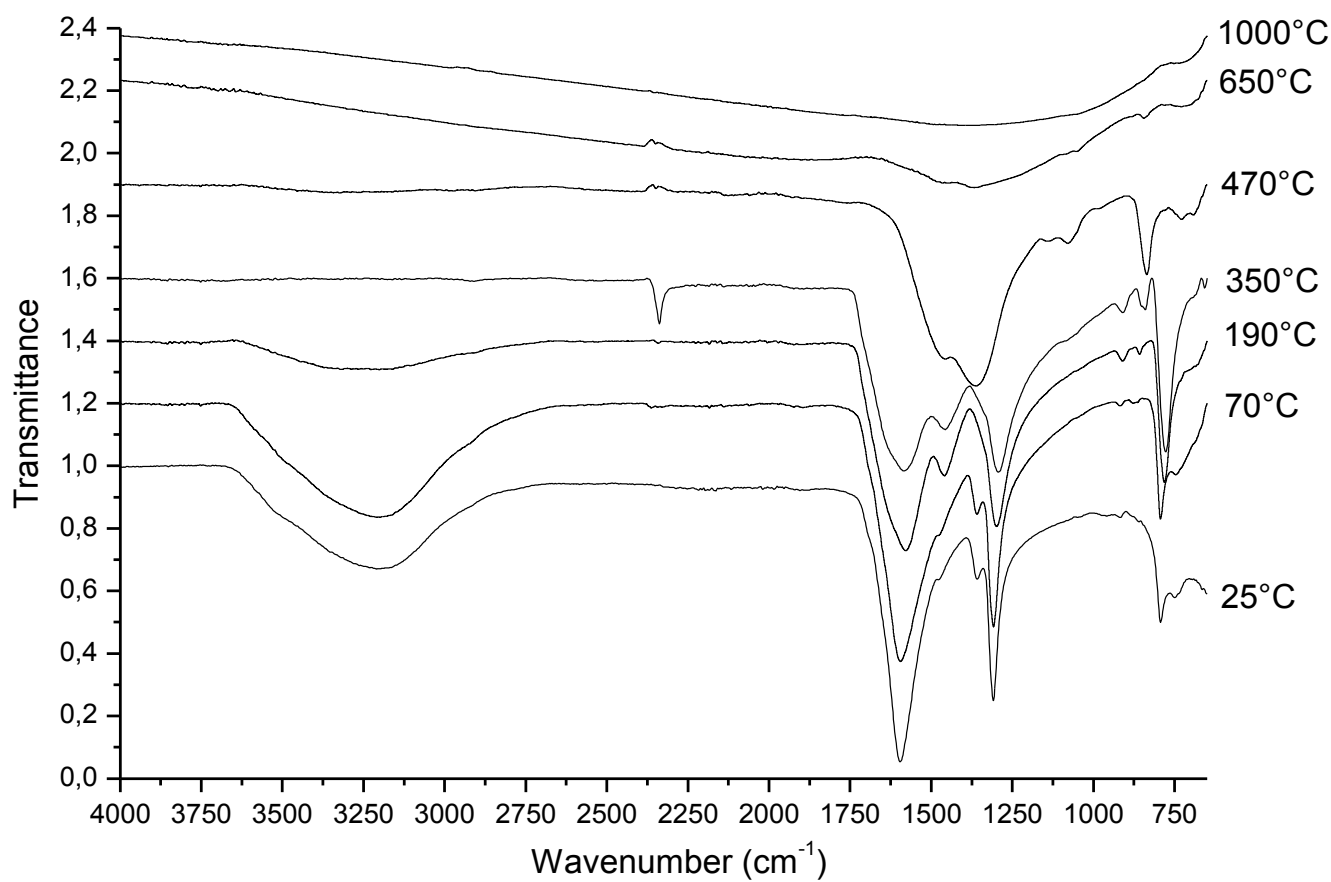

Figure 8 : FT-IR spectra of $Ce_2(C_2O_4)_3 \cdot nH_2O$ thermal decomposition intermediates under argon.

IR spectroscopy of intermediate products obtained from the calcination of cerium oxalate under inert atmosphere (Figure 8) confirms that the water absorption signal (broad band around 2700 – 3600 cm$^{-1}$) decreases first, while the oxalate group bands broaden (1585, 1460, 1350, 1300, 910, 860 and 790 cm$^{-1}$). At 350°C, the water signal has totally disappeared and characteristic bands of carbonate groups arise (1500, 1360, 1080 and 840 cm$^{-1}$). Moreover, sorbed $CO_2$ is observed within this intermediate, as illustrates the weak band at 2340 cm$^{-1}$, which is consistent with the beginning of the oxalate groups decomposition. Only carbonate bands can be seen on the spectrum of the compound calcined at 470°C, and a few traces remain at 650°C.

Furthermore, HT-XRD under both inert and oxidizing atmospheres (see supporting information) reveals that cerium oxalate hydrate becomes amorphous as soon as the dehydration starts, until cubic cerium oxide (JCPDS 00-034-0394) develops, around 425°C and 250°C respectively.

Based on the results from the different analysis techniques, the following mechanisms for the thermal decomposition of cerium oxalate under argon (i) and air (ii) are suggested:

(i) Under argon:

$$Ce_2(C_2O_4)_3 \cdot 8H_2O \xrightarrow{90-300°C} Ce_2(C_2O_4)_3 + 8\ H_2O$$

$$Ce_2(C_2O_4)_3 \xrightarrow{350-500°C} Ce_2O_2CO_3 + 2\ CO_2 + 3\ CO$$

$$2x\ CO_{(g)} \xrightarrow{450°C} x\ CO_{2(g)} + x\ C_{(s)}\ \text{(Boudouard reaction)}$$

$$Ce_2O_2CO_3 \xrightarrow{500-800°C} 2\ CeO_2 + CO$$

$$y\ CO_{2(g)} + y\ C_{(s)} \xrightarrow{550°C} 2y\ CO_{(g)}\ \text{(Boudouard reaction)}$$

(ii) In air:

$$Ce_2(C_2O_4)_3 \cdot 8H_2O \xrightarrow{90-220°C} Ce_2(C_2O_4)_3 + 8\ H_2O$$

$$Ce_2(C_2O_4)_3 \xrightarrow{270-450°C} 2\ CeO_2 + 2\ CO_2 + 4\ CO$$

$$CO + \tfrac{1}{2}\ O_2 \rightarrow CO_2$$

*1.3. Residual carbon content in the lanthanide oxides*

The residual carbon contents in the neodymium and cerium oxides obtained after calcination of the oxalate precursors at 1000°C under argon and in air were quantified by C-S analyser. Under oxidizing atmosphere, the carbon content was below the detection limit (i.e. 0,01w%), regardless of the heating rate. On the contrary, the oxides calcined under inert atmosphere retained a significant amount of elemental carbon: the results are presented in Table 6.

Table 6. Residual carbon content in the lanthanide oxides obtained after calcination at 1000°C under argon

| Compounds analyzed | Mass carbon content $\left(\dfrac{m_C}{m_{compound}}\right)$ | Molar carbon content $\left(\dfrac{n_C}{n_{oxide}}\right)$ |
|---|---|---|
| $Nd_2O_3$ calcined at 2°C/min | 1.69 % | 0.42 |
| $Nd_2O_3$ calcined at 5°C/min | 1.60 % | 0.42 |
| $Nd_2O_3$ calcined at 10°C/min | 1.55 % | 0.42 |
| $CeO_2$ calcined at 2°C/min | 1.57 % | 0.23 |
| $CeO_2$ calcined at 5°C/min | 1.46 % | 0.21 |
| $CeO_2$ calcined at 10°C/min | 1.37 % | 0.20 |
| $CeO_2$ calcined at 10°C/min, larger volume | 1.03 % | 0.15 |

Elemental carbon is formed during the second phase of the thermal decomposition, i.e. as soon as the oxalate groups start to decompose. Under air, carbon dioxide is the main gas released, and potentially evolved carbon monoxide is immediately oxidized by air, according to the reaction: $CO + \tfrac{1}{2}\ O_2 \rightarrow CO_2$. On the contrary, calcination under inert atmosphere leads to the substantial release of carbon monoxide, which undergoes partial disproportionation to give carbon dioxide and elemental carbon: $2\ CO \rightarrow C + CO_2$ (Boudouard reaction). At higher temperature, reverse Boudouard reaction leads to the consumption of elemental carbon and $CO_2$ to re-form CO.

Thus, under argon, some elemental carbon remains in the oxide even after calcination up to 1000°C. The mean residual carbon content in these oxides is about 0.19 mole of $C_{(s)}$ per mole of $CeO_2$ and 0.42 mole of $C_{(s)}$ per mole of $Nd_2O_3$. These values tend to decrease when the heating rate increases.

Moreover, an experiment was carried out with a more important quantity of the cerium oxalate precursor (around 2.0 g). It revealed that the carbon content in the residual oxide was less than for the calcination of 20-30 mg of oxalate at the same heating rate. The black colour due to elemental carbon was localised on the periphery of the powder in the crucible, while the bulk of the sample was light yellow (colour of pure cerium oxide). This is an additional proof that during the diffusion of evolved gases toward the surface of the powder, carbon monoxide disproportionate to form elemental carbon.

## 2. Thermodynamics and kinetics of the thermal decomposition

*2.1. Thermodynamic studies*

In order to determine the thermodynamic and kinetic parameters describing the thermal decomposition of neodymium and cerium oxalates, the thermogravimetric studies were carried out at three different temperatures under each atmosphere, following the method of previous investigators like Bigda.[13]

The peaks of the DTA or DSC curves were integrated to calculate the heats of reaction corresponding to each thermal decomposition step. The results are presented in Table 7 for neodymium oxalate and Table 8 for cerium oxalate. No data was available in the literature for comparison. As the enthalpy of a given reaction is by definition constant regardless of the heating rate, the small differences observed between the calculated values for each heating rate can be attributed to the uncertainty of the analyses.

Table 7. Heats of reaction corresponding to the calcination of neodymium oxalate under argon and in air

| Atm. | Reaction step | 2°C/min | 5°C/min | 10°C/min |
|---|---|---|---|---|
| Argon | $Nd_2(C_2O_4)_3.8H_2O \rightarrow Nd_2(C_2O_4)_3.2H_2O$ | $\Delta H_1 = 222.8 \pm 4.0$ kJ/mol | $\Delta H_1 = 226.9 \pm 4.0$ kJ/mol | $\Delta H_1 = 229.6 \pm 4.0$ kJ/mol |
| | $Nd_2(C_2O_4)_3.2H_2O \rightarrow Nd_2(C_2O_4)_3$ | $\Delta H_2 = 33.1 \pm 2.0$ kJ/mol | $\Delta H_2 = 34.1 \pm 2.0$ kJ/mol | $\Delta H_2 = 31.5 \pm 2.0$ kJ/mol |
| | $Nd_2(C_2O_4)_3 \rightarrow Nd_2(CO_3)_3$ | $\Delta H_3 = 52.3 \pm 2.0$ kJ/mol | $\Delta H_3 = 53.0 \pm 2.0$ kJ/mol | $\Delta H_3 = 54.0 \pm 2.0$ kJ/mol |
| | $Nd_2(CO_3)_3 \rightarrow Nd_2O_2CO_3$ | / | $\Delta H_4 = 5.5 \pm 1.0$ kJ/mol | $\Delta H_4 = 4.7 \pm 1.0$ kJ/mol |
| | $Nd_2O_2CO_3 \rightarrow Nd_2O_3$ | / | $\Delta H_5 = 14.9 \pm 1.0$ kJ/mol | $\Delta H_5 = 13.9 \pm 1.0$ kJ/mol |
| Air | $Nd_2(C_2O_4)_3.8H_2O \rightarrow Nd_2(C_2O_4)_3.2H_2O$ | $\Delta H_1 = 228.0 \pm 3.0$ kJ/mol | $\Delta H_1 = 230.3 \pm 3.0$ kJ/mol | $\Delta H_1 = 225.2 \pm 3.0$ kJ/mol |
| | $Nd_2(C_2O_4)_3.2H_2O \rightarrow Nd_2(C_2O_4)_3$ | $\Delta H_2 = 35.5 \pm 2.0$ kJ/mol | $\Delta H_2 = 34.6 \pm 2.0$ kJ/mol | $\Delta H_2 = 36.2 \pm 2.0$ kJ/mol |
| | $Nd_2(C_2O_4)_3 \rightarrow Nd_2(C_2O_4)_{5/2}(CO_3)_{1/2}$ | $\Delta H_3 = -10.9 \pm 1.0$ kJ/mol | $\Delta H_3 = -10.4 \pm 1.0$ kJ/mol | $\Delta H_3 = -11.1 \pm 1.0$ kJ/mol |
| | $Nd_2(C_2O_4)_{5/2}(CO_3)_{1/2} \rightarrow Nd_2(CO_3)_3$ | $\Delta H_4 = -16.2 \pm 2.0$ kJ/mol | $\Delta H_4 = -18.2 \pm 2.0$ kJ/mol | $\Delta H_4 = -15.8 \pm 2.0$ kJ/mol |
| | $Nd_2(CO_3)_3 \rightarrow Nd_2O_2CO_3$ | $\Delta H_5 = -157.5 \pm 3.0$ kJ/mol | $\Delta H_5 = -159.5 \pm 3.0$ kJ/mol | $\Delta H_5 = -165.5 \pm 3.0$ kJ/mol |
| | $Nd_2O_2CO_3 \rightarrow Nd_2O_3$ | $\Delta H_6 = 32.6 \pm 2.0$ kJ/mol | $\Delta H_6 = 32.4 \pm 2.0$ kJ/mol | $\Delta H_6 = 32.1 \pm 2.0$ kJ/mol |

Table 8. Heats of reaction corresponding to the calcination of cerium oxalate under argon and in air

| Atm. | Reaction step | 2°C/min | 5°C/min | 10°C/min |
|---|---|---|---|---|
| Argon | $Ce_2(C_2O_4)_3.8H_2O \rightarrow Ce_2(C_2O_4)_3$ | $\Delta H_1 = 397.5 \pm 4.0$ kJ/mol | $\Delta H_1 = 410.0 \pm 4.0$ kJ/mol | $\Delta H_1 = 409.5 \pm 4.0$ kJ/mol |
| | $Ce_2(C_2O_4)_3 \rightarrow Ce_2O_2CO_3$ | $\Delta H_2 = 77.1 \pm 2.0$ kJ/mol | $\Delta H_2 = 73.0 \pm 2.0$ kJ/mol | $\Delta H_2 = 74.5 \pm 2.0$ kJ/mol |
| | $Ce_2O_2CO_3 \rightarrow 2\,CeO_2$ | $\Delta H_3 = 3.8 \pm 1.0$ kJ/mol | $\Delta H_3 = 6.5 \pm 1.0$ kJ/mol | $\Delta H_3 = 7.9 \pm 1.0$ kJ/mol |
| Air | $Ce_2(C_2O_4)_3.8H_2O \rightarrow Ce_2(C_2O_4)_3$ | $\Delta H_1 = 379 \pm 8$ kJ/mol | $\Delta H_1 = 390 \pm 4$ kJ/mol | $\Delta H_1 = 389 \pm 4$ kJ/mol |
| | $Ce_2(C_2O_4)_3 \rightarrow 2\,CeO_2$ | $\Delta H_2 = -539 \pm 10$ kJ/mol | $\Delta H_2 = -555 \pm 5$ kJ/mol | $\Delta H_2 = -552 \pm 5$ kJ/mol |

*2.2. Kinetic studies*

For a thermal decomposition reaction: $aA_{(solid)} \xrightarrow{\Delta T} bB_{(solid)} + cC_{(gas)}$, the reaction rate can be expressed as the disappearance of reagent A: $v = \frac{d\alpha}{dt} = k(T).f(\alpha)$. The conversion degree $\alpha$ is defined as: $\alpha = \frac{m_0 - m_t}{m_0 - m_\infty}$ with $m_0$, $m_t$ et $m_\infty$ the initial mass of the sample, the mass at time $t$ and the final mass respectively. The rate constant $k$ is assumed to follow the Arrhenius law: $k(T) = A\exp(-E/RT)$, with $A$ being the pre-exponential factor, $E$ the activation energy and $R$ the gas constant. $f(\alpha)$ is a mathematic function describing the kinetic model of the reaction. Numerous kinetic models have been proposed in the literature, and the models used in this study are shown in Table 9.

For non-isothermal kinetics, the variation in the conversion degree is measured with respect to time, while a linear heating rate $\beta$ is imposed: $T = T_0 + \beta t$. Thus: $\frac{d\alpha}{dt} = \beta \frac{d\alpha}{dT}$.

The isoconversional method compares the data obtained for a given $\alpha$ at different heating rates: $ln\left(\frac{d\alpha}{dt}\right) = ln\left(\beta \frac{d\alpha}{dT}\right) = ln(A.f(\alpha)) - \frac{E}{RT}$. Thus, the determination of the kinetic model function is not needed to calculate $E$, but a function $f(\alpha)$ must be postulated to calculate $A$.

Table 9. Kinetic model functions of the Netzsch Thermokinetics 3.1 software [14]

| Code | f($\alpha$) | Reaction type |
|---|---|---|
| F1 | $(1-\alpha)$ | First-order reaction |
| F2 | $(1-\alpha)^2$ | Second-order reaction |
| Fn | $(1-\alpha)^n$ | $n^{th}$-order reaction |
| R2 | $2(1-\alpha)^{1/2}$ | Two-dimensional phase boundary reaction |
| R3 | $3(1-\alpha)^{2/3}$ | Three-dimensional phase boundary reaction |
| D1 | $1/2\alpha$ | One-dimensional diffusion |
| D2 | $-1/ln(1-\alpha)$ | Two-dimensional diffusion |
| D3 | $(3/2)(1-\alpha)^{1/3}/[(1-\alpha)^{-1/3}-1]$ | Three-dimensional diffusion (Jander's type) |
| D4 | $(3/2)/[(1-\alpha)^{-1/3}-1]$ | Three-dimensional diffusion (Ginstling-Brounstein's type) |
| B1 | $\alpha(1-\alpha)$ | Simple Prout-Tompkins equation |
| Bna | $(1-\alpha)^n \alpha^a$ | Expanded Prout-Tompkins equation (na) |
| C1 P | $(1-\alpha)(1+\alpha K_{cat})$ | First-order reaction with autocatalysis through the product P |
| Cn P | $(1-\alpha)^n(1+\alpha K_{cat})$ | $n^{th}$-order reaction with autocatalysis through the product P |
| A2 | $2(1-\alpha)[-ln(1-\alpha)]^{1/2}$ | Two-dimensional nucleation |
| A3 | $3(1-\alpha)[-ln(1-\alpha)]^{2/3}$ | Three-dimensional nucleation |
| An | $n(1-\alpha)[-ln(1-\alpha)]^{(n-1)/n}$ | $n$-dimensional nucleation/nucleus growth according to Avrami-Erofeev |

The kinetics of the thermal decomposition of the two lanthanide oxalates were examined using the *Netzsch Thermokinetics 3.1* software package [14]. This software allows calculation of the kinetic parameters from the experimental TG or DTA data obtained for at least three different heating rates under a same atmosphere. First, a model-free isoconversional analysis (Friedman or Ozawa-Flynn-Wall methods) is applied to the non-isothermal measurements to estimate the activation energy and the pre-exponential factor as a function of the conversion degree. Then those values are used as starting parameters for a multivariate non-linear regression in order to fit the TG data to multi-step mechanisms, allowing the determination of the kinetic model corresponding to each elementary step and refinement of the kinetic parameter values.

Tables 10 and 11 summarise the kinetic parameters determined for the dehydration steps of neodymium and cerium oxalates and compare them with the results previously reported in the literature. The TG curves were best fit with an "n-th order with autocatalysis" kinetic model: $f(\alpha) = (1-\alpha)^n(1+K_{cat}\alpha)$. Even though the results found in the literature sometimes differ from a factor 3, the kinetic parameters calculated in this study are consistent with the order of magnitude of the previously reported data, obtained using different kinetic models.

Table 10. Kinetic parameters for the calcination of neodymium oxalate under argon and in air

| Atm. | Step | This work | | | | Literature | | |
|---|---|---|---|---|---|---|---|---|
| | | log A (s$^{-1}$) | E (kJ/mol) | log K$_{cat}$ | n | E (kJ/mol) | n | Ref. |
| **Argon** | 1 | 7.3 | 76.3 | - 4.0 | 2.0 | 89.4 | 2 | Under N$_2$ [5g] |
| | 2 | 14.9 | 161.7 | - 4.0 | 3.0 | / | / | / |
| | 3 | / | 323.3 | / | / | / | / | / |
| | 4 | / | 553.5 | / | / | / | / | / |
| | 5 | 5.5 | 153.4 | - 4.0 | 1.8 | / | / | / |
| **Air** | 1 | 5.2 | 59.3 | - 4.0 | 1.2 | 57 | 0.8 | [5b] |
| | | | | | | 50.2 | / | [5b] |
| | | | | | | 141.7 | 1 | [5f] |
| | 2 | 5.6 | 79.6 | 1.1 | 2.3 | 126.1 | 1 | [5f] |
| | 3 | / | 407.1 | / | / | 800.5 | 3 | [5f] |
| | 4 | / | 247.3 | / | / | | | |
| | 5 | 14.1 | 241.3 | - 4.0 | 2.7 | / | / | / |
| | 6 | 12.2 | 286.1 | 2.0 | 1.1 | 663.2 | 3 | [5f] |

Table 11. Kinetic parameters for the calcination of cerium oxalate under argon and in air

| Atm. | Step | This work | | | | Literature | | |
|---|---|---|---|---|---|---|---|---|
| | | log A (s$^{-1}$) | E (kJ/mol) | log K$_{cat}$ | n | E (kJ/mol) | n | Ref. |
| **Argon** | 1 | 5.7 | 66.8 | 0.5 | 1.7 | / | / | / |
| | 2 | / | 197.9 | / | / | / | / | / |
| | 3 | / | 646.6 | / | / | / | / | / |
| **Air** | 1 | 5.9 | 66.0 | 0.2 | 1.6 | 83.7 | 0.9 | [6a] |
| | | | | | | 53.2 | 0.6 | [5b] |
| | | | | | | 37.3 | 0.6 | [5b] |
| | | | | | | 78.2 | / | [6f] |
| | | | | | | 151.9 | 1 | [5f] |
| | 2 | 10.9 | 155.1 | -0.4 | 1.1 | 215.6 | 1 | [6a] |
| | | | | | | 112.6 | / | [6f] |
| | | | | | | 728.1 | 3 | [5f] |

An example of a TG curve fitted with the calculated kinetic parameters is given in Figure 9.

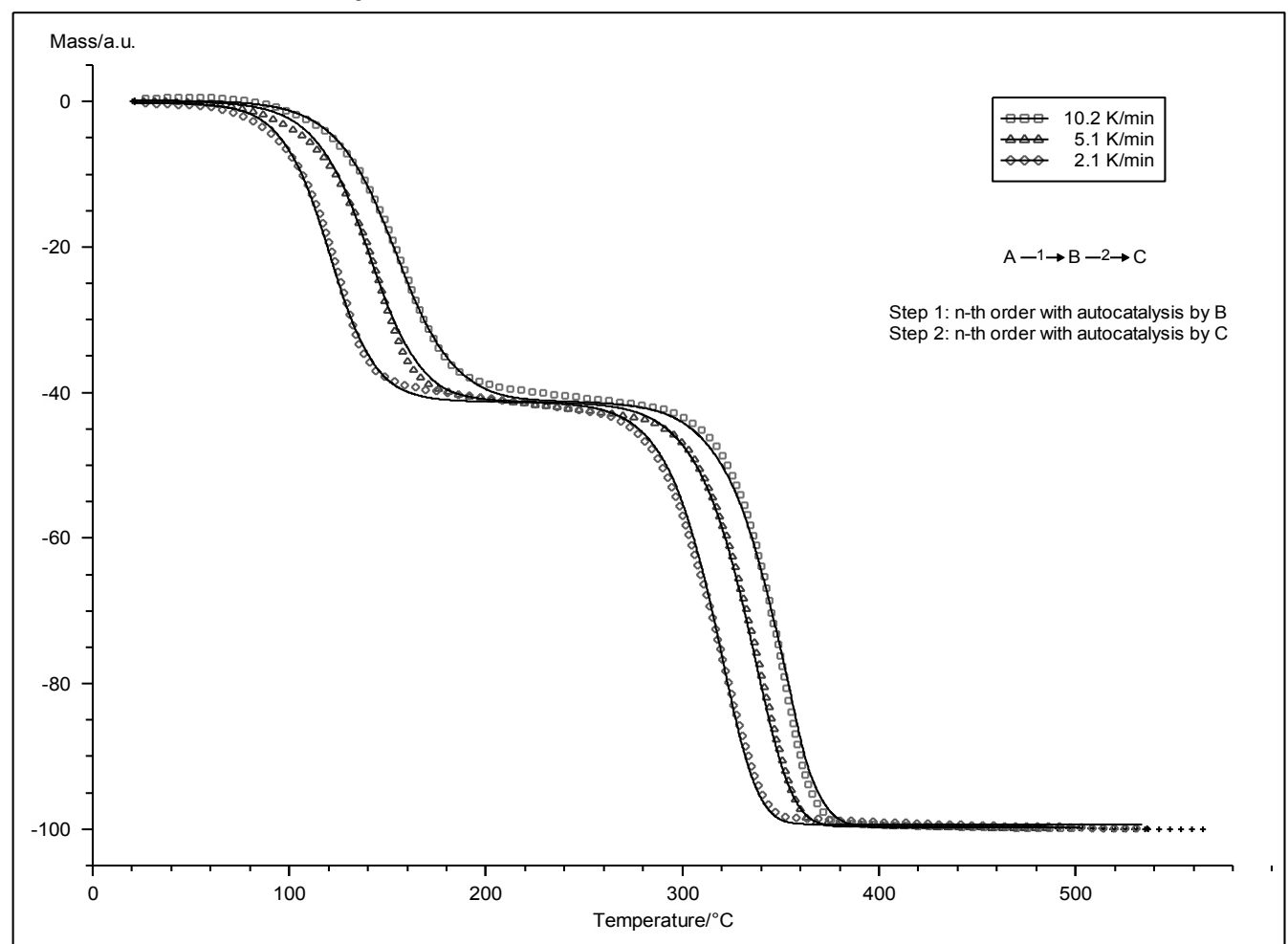

Figure 9 : Fit of the TG curve for the thermal decomposition of cerium oxalate in air with the calculated kinetic parameters.

The kinetic parameters calculated for each step are "apparent parameters", accounting for the different phenomena that superimpose within the corresponding temperature range. Therefore, when concurrent reactions occur within the same temperature range, the calculation of the kinetic parameters is hampered, leading to the fit of the TG curves with reaction orders greater than 3.0, which has no physical meaning. In this case, the calculated kinetic parameters weren't indicated, except for the activation energy which was reliably obtained by model-free isoconversional analysis.

The fits of the TG curves correlate strongly with the calculated kinetic parameters for each system ($R^2 >$ 0.9998) and the values of the activation energy are in good agreement with the corresponding results previously published in the literature, obtained using other kinetic methods. Nevertheless, these results needs to be put into perspective as there exists to date no independent and reliable way to establish solid-state kinetic parameters, which are dependent to some degree on the kinetic method employed.

To supplement the kinetic studies, microscopic observations of the powder evolution with increasing temperature were made with SEM. The results obtained for the thermal decomposition of neodymium oxalate under argon are shown in Figure 10, and for cerium oxalate in Figure 11.

In both cases, the rod-like morphology of the compounds is preserved during the thermal decomposition. When decomposition of the oxalate groups begins, blisters and open porosity are observed on the surface of the crystallites, created by gases escaping from the bulk. The oxides display rough faces due to increased porosity and cracking, or even maybe to residual elemental carbon deposits. These observations support the fact that solid-gas interactions, which are not translated on the mass loss curves, prevent the calculation of physically meaningful kinetic parameters.

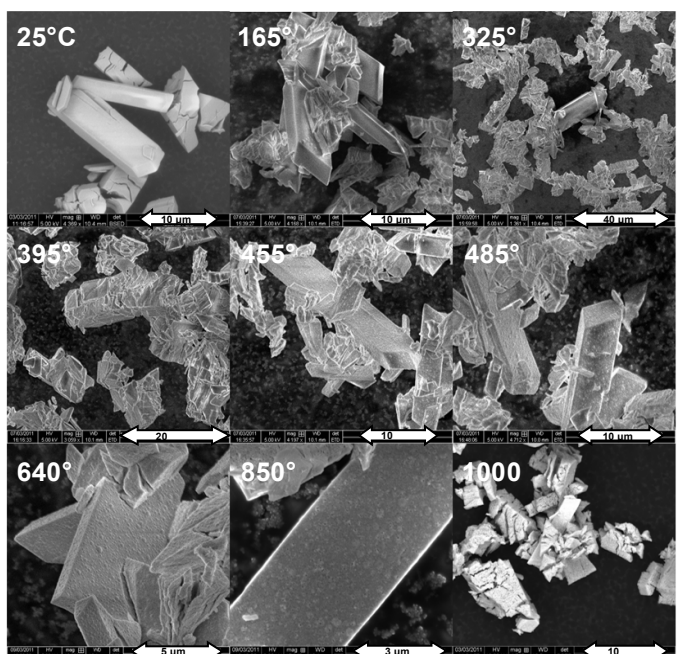

Figure 10 : SEM observations of the thermal decomposition of neodymium oxalate under argon.

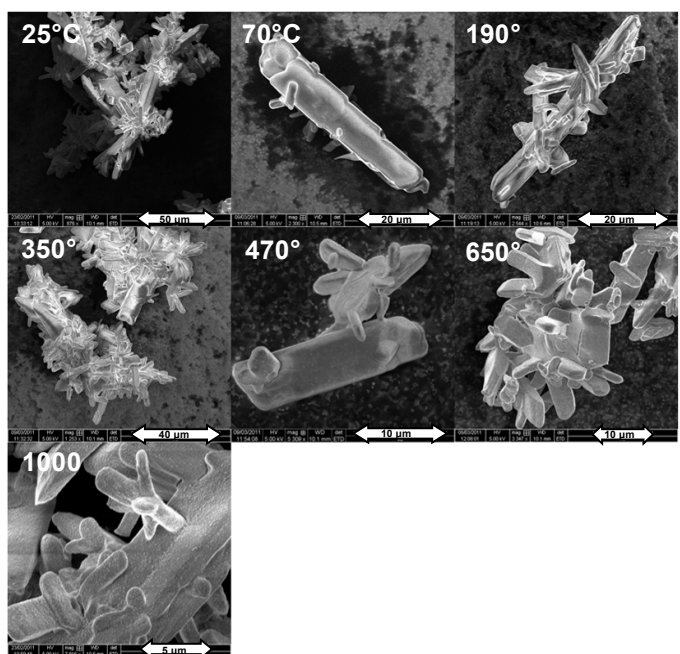

Figure 11 : SEM observations of the thermal decomposition of cerium oxalate under argon.

## 3. Comparison with the thermal decomposition of actinide (III) oxalates: example of Pu(III) oxalate

### 3.1. Thermal decomposition of plutonium(III) oxalate

The thermal decomposition of plutonium(III) oxalate was studied as a representative of the actinide(III) behaviour. Plutonium oxalate was synthesised as stated in the experimental section, and used as freshly prepared to prevent the compound from deteriorating by self-radiolysis and/or self-induced oxidation.

To verify that the compound synthesized was the expect plutonium(III) oxalate, it was characterised by different techniques (see supporting information).

The UV-vis spectra displayed the characteristic absorption peaks of trivalent plutonium, in the region of 500-700 nm, and the principal absorption peak for tetravalent plutonium at 475 nm (intense) was not observed [15], ensuring that the oxidation state of plutonium within the precipitate was trivalent.

The functional groups of the precipitate were controlled by FT-IR spectroscopy, showing the characteristic vibration bands of hydrated oxalate, with the large signal between 2700 and 3600 cm$^{-1}$ corresponding to the water molecules and the remaining peaks attributable to the oxalate group vibrations (1620, 1470, 1360, 1320, 920 and 800 cm$^{-1}$).

The synthesised plutonium oxalate was then analyzed by X-ray diffraction, as well as the residual compound after calcination at 700°C. The resulting spectra were compared to the JCPDS database and matched respectively with monoclinic plutonium(III) oxalate hydrate (JCPDS 00-020-0842) and cubic plutonium(IV) oxide (JCPDS 00-041-1170).

For the thermogravimetric studies, samples of about 45 mg of freshly synthesised plutonium(III) oxalate were calcined under argon and in air at 2°C/min (Figures 12 and 13).

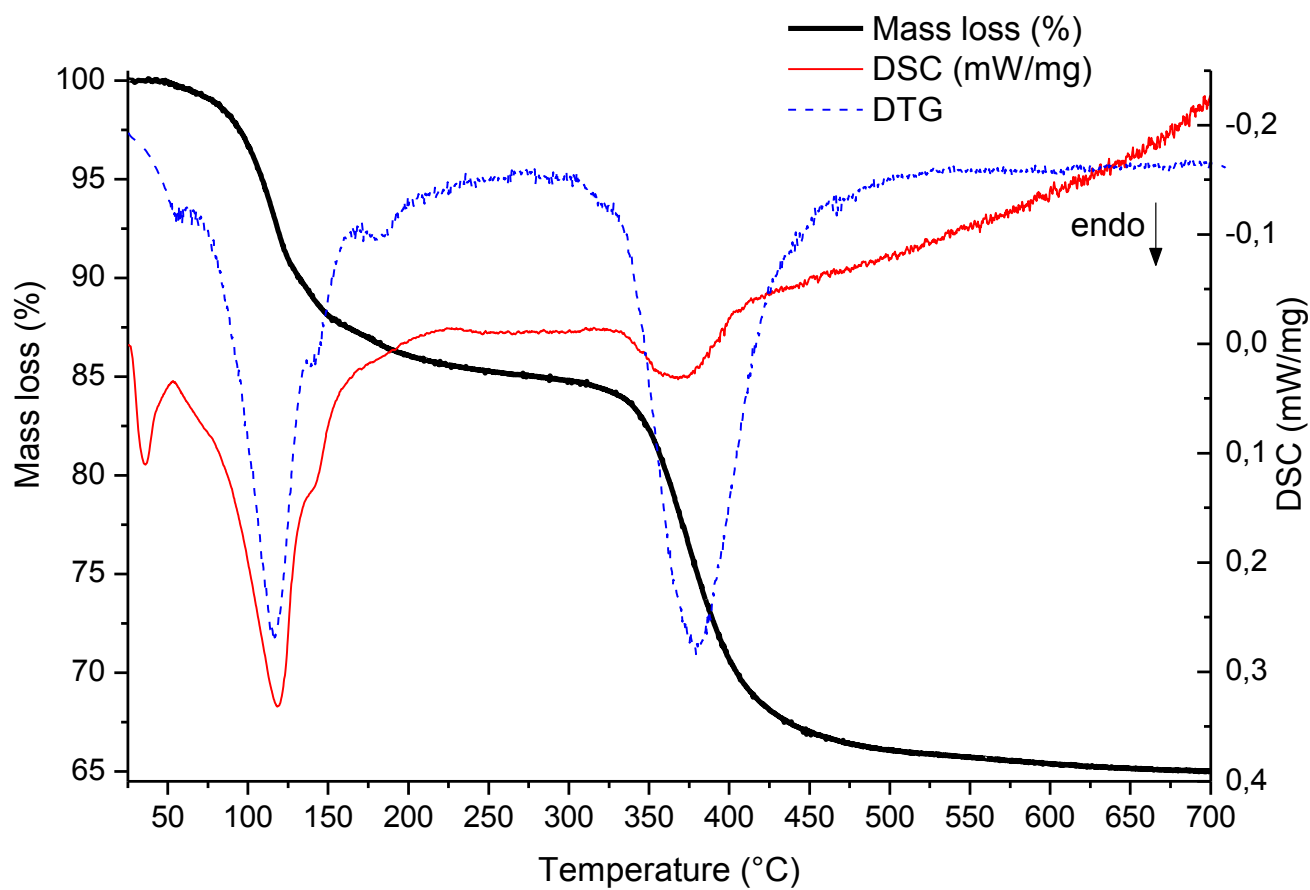

Figure 12 : Thermal decomposition of plutonium oxalate under streaming argon (3 NL/h) on heating at 2°C/min.

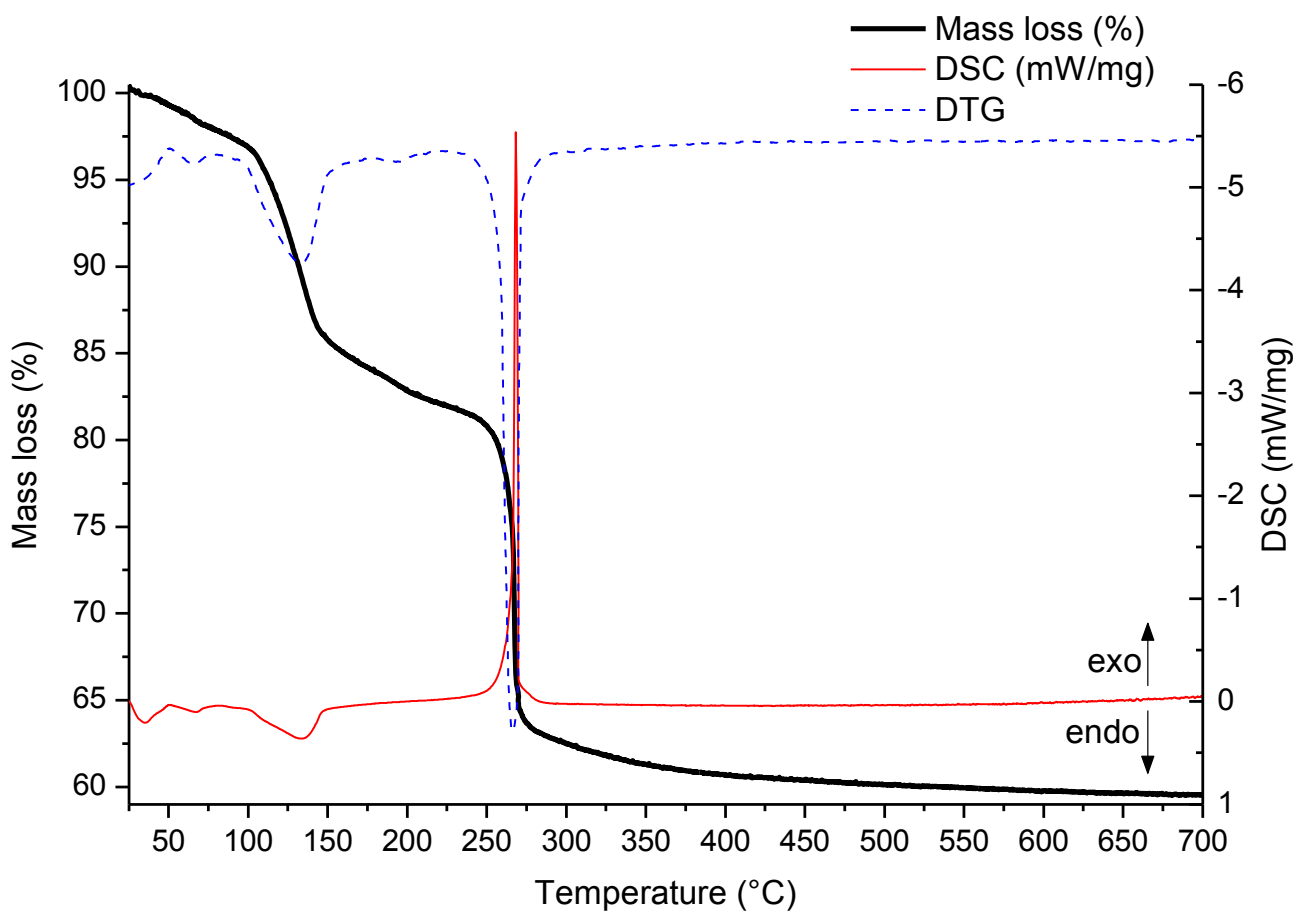

Figure 13 : Thermal decomposition of plutonium oxalate under static air on heating at 2°C/min.

When normalising the TG curves with respect to the oxide, it appeared that the hydration state of the starting oxalate was not the same under air and under argon. This is due to the high vacuum used to remove traces of air in the thermobalance before working under inert argon atmosphere.

The peaks of the DTG curve (first derivative of the TG curve) define the consecutive steps of the thermal decomposition mechanism. Thus, it appears that the decomposition of plutonium(III) oxalate occurs in four steps under argon and five in air. As usual, the dehydration steps are endothermic regardless of the atmosphere and the oxalate decomposition step is strongly exothermic in air while it is endothermic under argon.

Tables 12 and 13 compare experimental and theoretical mass losses, corresponding to the TG analyses of plutonium(III) oxalate under argon and in air.

Table 12. Comparison between the experimental and theoretical mass losses for the calcination of plutonium(III) oxalate under argon at 2°C/min

| | Proposed reaction step | Temperature range (°C) | Theoretical mass loss (%) | Experimental mass loss (%) |
|---|---|---|---|---|
| 1 | $Pu_2(C_2O_4)_3 \cdot 7H_2O \rightarrow Pu_2(C_2O_4)_3 \cdot 2H_2O$ | 50 – 135 | 10.4 | 10.2 |
| 2 | $Pu_2(C_2O_4)_3 \cdot 2H_2O \rightarrow Pu_2(C_2O_4)_3 \cdot H_2O$ | 135 – 160 | 2.1 | 2.2 |
| 3 | $Pu_2(C_2O_4)_3 \cdot H_2O \rightarrow Pu_2(C_2O_4)_3$ | 160 – 275 | 2.1 | 2.5 |
| 4 | $Pu_2(C_2O_4)_3 \rightarrow 2\ PuO_2$ | 275 – 700 | 23.0 | 20.1 |

Table 13. Comparison between the experimental and theoretical mass losses for the calcination of plutonium(III) oxalate in air at 2°C/min

| | Proposed reaction step | Temperature range (°C) | Theoretical mass loss (%) | Experimental mass loss (%) |
|---|---|---|---|---|
| 1 | $Pu_2(C_2O_4)_3 \cdot 9H_2O \rightarrow Pu_2(C_2O_4)_3 \cdot 8H_2O$ | 30 – 80 | 2.0 | 2.1 |
| 2 | $Pu_2(C_2O_4)_3 \cdot 8H_2O \rightarrow Pu_2(C_2O_4)_3 \cdot 2H_2O$ | 80 – 140 | 11.8 | 11.9 |
| 3 | $Pu_2(C_2O_4)_3 \cdot 2H_2O \rightarrow Pu_2(C_2O_4)_3 \cdot H_2O$ | 140 – 180 | 1.9 | 2.1 |
| 4 | $Pu_2(C_2O_4)_3 \cdot H_2O \rightarrow Pu_2(C_2O_4)_3$ | 180 – 225 | 2.0 | 2.1 |
| 5 | $Pu_2(C_2O_4)_3 \rightarrow 2\ PuO_2$ | 225 – 400 | 22.1 | 22.3 |

The experimental and theoretical mass losses are in good agreement. It appeared that even without vacuum, the starting oxalate was a nonahydrate, thus the first dehydration step under static air accounts for the loss of one water molecule. On the other hand, the first dehydration step under argon accounts only for five water molecules, and the experimental mass loss for the oxalate decomposition step is smaller than expected because of the residual carbon present in the oxide at the end of the calcination [7d]. Superimposing the TG curves of the thermal decomposition of plutonium oxalate in air and under argon (cf. Figure 14), it appears that the gap between the mass losses at 700°C is 2.5%: this is attributable to the presence of elemental carbon (or possible oxalate residue), as the decomposition is not complete at this temperature.

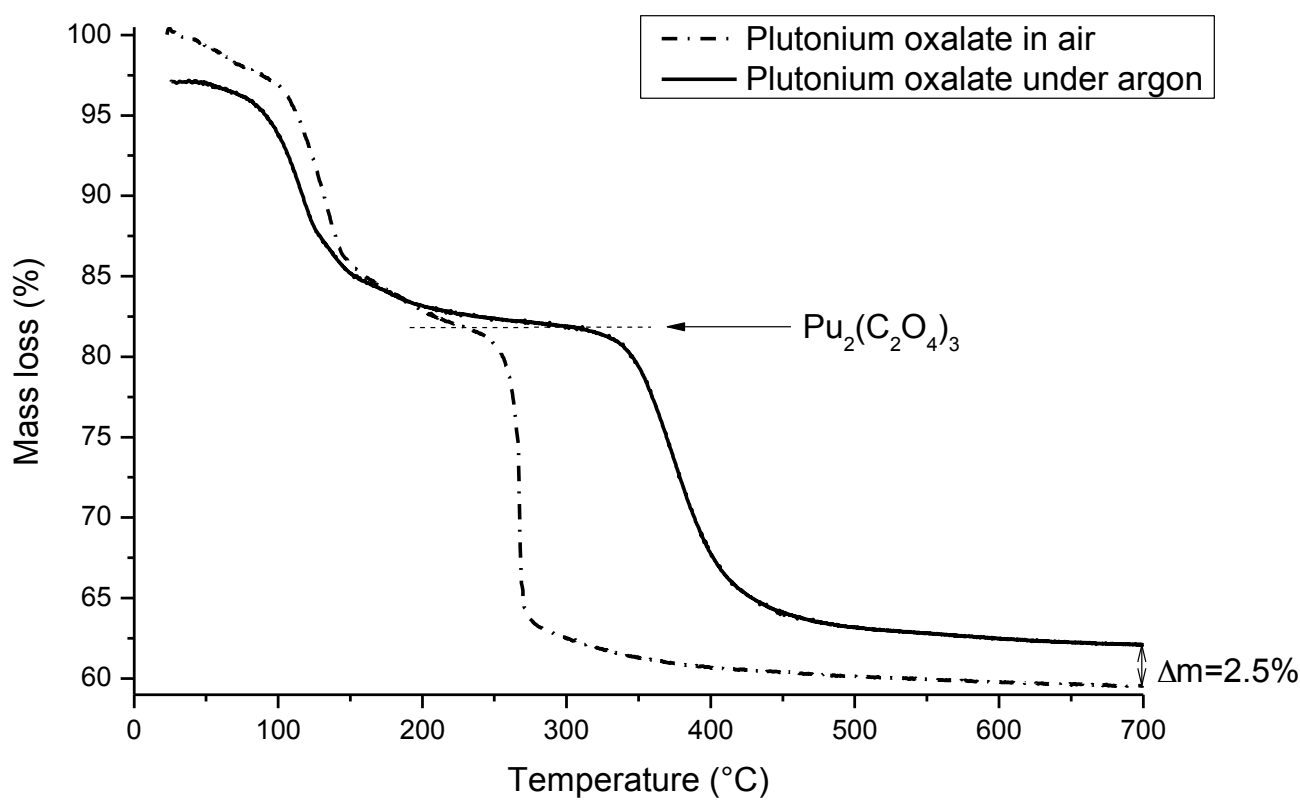

Figure 14 : Comparison between the thermal decomposition of plutonium oxalate in air and under argon.

Thermodynamic data for the thermal decomposition of plutonium(III) oxalate under argon and in air were calculated by integration of the DSC peaks. The results are reported in Table 14, along with the only reference from the literature, from 1969, where the authors calculated the heat of reaction from the equation described by Spiels *et al.* (in 1945). The discrepancy between the results can either be explained by the differing calculation methods, or their starting oxalate was more dehydrated than the one used in this study.

Table 14. Heats of reaction corresponding to the calcination of plutonium oxalate under argon and in air

| Atm. | Reaction step | ΔH (kJ/mol) | | |
|---|---|---|---|---|
| | | This study | Literature | Ref. |
| **Argon** | $Pu_2(C_2O_4)_3 \cdot 7H_2O \rightarrow Pu_2(C_2O_4)_3$ | 368.5 | / | |
| | $Pu_2(C_2O_4)_3 \rightarrow 2\ PuO_2$ | 79.8 | / | |
| **Air** | $Pu_2(C_2O_4)_3 \cdot 9H_2O \rightarrow Pu_2(C_2O_4)_3 \cdot 8H_2O$ | 22.6 | 180.0 | [7b] |
| | $Pu_2(C_2O_4)_3 \cdot 8H_2O \rightarrow Pu_2(C_2O_4)_3$ | 301.0 | for complete dehydration | |
| | $Pu_2(C_2O_4)_3 \rightarrow 2\ PuO_2$ | − 454.0 | − 293.0 | [7b] |

Based on the results from the different analyses, the following mechanisms for the thermal decomposition of plutonium oxalate under argon (i) and air (ii) are proposed:

(i) Under argon:

$Pu_2(C_2O_4)_3 \cdot 7H_2O \xrightarrow{80-135°C} Pu_2(C_2O_4)_3 \cdot 2H_2O + 5\ H_2O$

$Pu_2(C_2O_4)_3 \cdot 2H_2O \xrightarrow{135-160°C} Pu_2(C_2O_4)_3 \cdot H_2O + H_2O$

$Pu_2(C_2O_4)_3 \cdot H_2O \xrightarrow{160-275°C} Pu_2(C_2O_4)_3 + H_2O$

$Pu^{III}_2(C_2O_4)_3 \xrightarrow{275-700°C} 2\ Pu^{IV}O_2 + 2\ CO_2 + 4\ CO$

$2\ CO \longrightarrow CO_2 + C$

(ii) In air:

$Pu_2(C_2O_4)_3 \cdot 9H_2O \xrightarrow{30-80°C} Pu_2(C_2O_4)_3 \cdot 8H_2O + H_2O$

$Pu_2(C_2O_4)_3 \cdot 8H_2O \xrightarrow{80-140°C} Pu_2(C_2O_4)_3 \cdot 2H_2O + 6\ H_2O$

$Pu_2(C_2O_4)_3 \cdot 2H_2O \xrightarrow{140-180°C} Pu_2(C_2O_4)_3 \cdot H_2O + H_2O$

$Pu_2(C_2O_4)_3 \cdot H_2O \xrightarrow{180-225°C} Pu_2(C_2O_4)_3 + H_2O$

$Pu_2(C_2O_4)_3 \xrightarrow{225-400°C} 2\ Pu^{IV}O_2 + 2\ CO_2 + 4\ CO$

$CO + \tfrac{1}{2} O_2 \longrightarrow CO_2$

The decomposition mechanism in air is consistent with the work of Subramanian [7b] and Sali [7d], as they both consider that the dehydrated oxalate is obtained prior to the decomposition of the oxalate groups, unlike Rao [7a] and Kozlova [7c] who proposed a monohydrated intermediate instead. As for the thermal decomposition mechanism under argon, it is in agreement with the one proposed by Rao [7a].

*3.2. Comparison between the thermal decomposition of lanthanide(III) and actinide(III) oxalates*

Lanthanide(III) and actinide(III) oxalate hydrates are isomorphic, but their structural arrangement being the same does not mean that their thermal decompositions follow exactly the same mechanism. Nevertheless, a parallelism can be found between the decomposition paths of lanthanide and actinide oxalates, especially between plutonium and cerium oxalate.

First of all, dehydration occurs in four steps for plutonium(III) oxalate hydrate, whereas it happens in one or two steps for cerium and neodymium oxalate respectively. However, the dihydrate intermediate is isolable for both neodymium and plutonium(III) oxalates.

Then, decomposition of the oxalate groups occurs in one step regardless of the atmosphere for plutonium(III) oxalate, as for cerium oxalate under air, while it involves the formation of one or several intermediates (carbonate mainly) for neodymium oxalate under both atmospheres and for cerium oxalate under argon.

Moreover, the decomposition of plutonium(III) and cerium oxalates in air is strongly exothermic. For both compounds, the metal cation undergoes oxidation during the last step of the thermal decomposition, as the oxalate in the trivalent oxidation state becomes the oxide $M^{+IV}O_2$. The formation of the resulting oxide is thus accelerated under oxidising atmosphere and is complete below 450°C. Conversely, the absence of redox properties for neodymium prevents the oxalate decomposition from being accelerated in air as compared to under argon atmosphere.

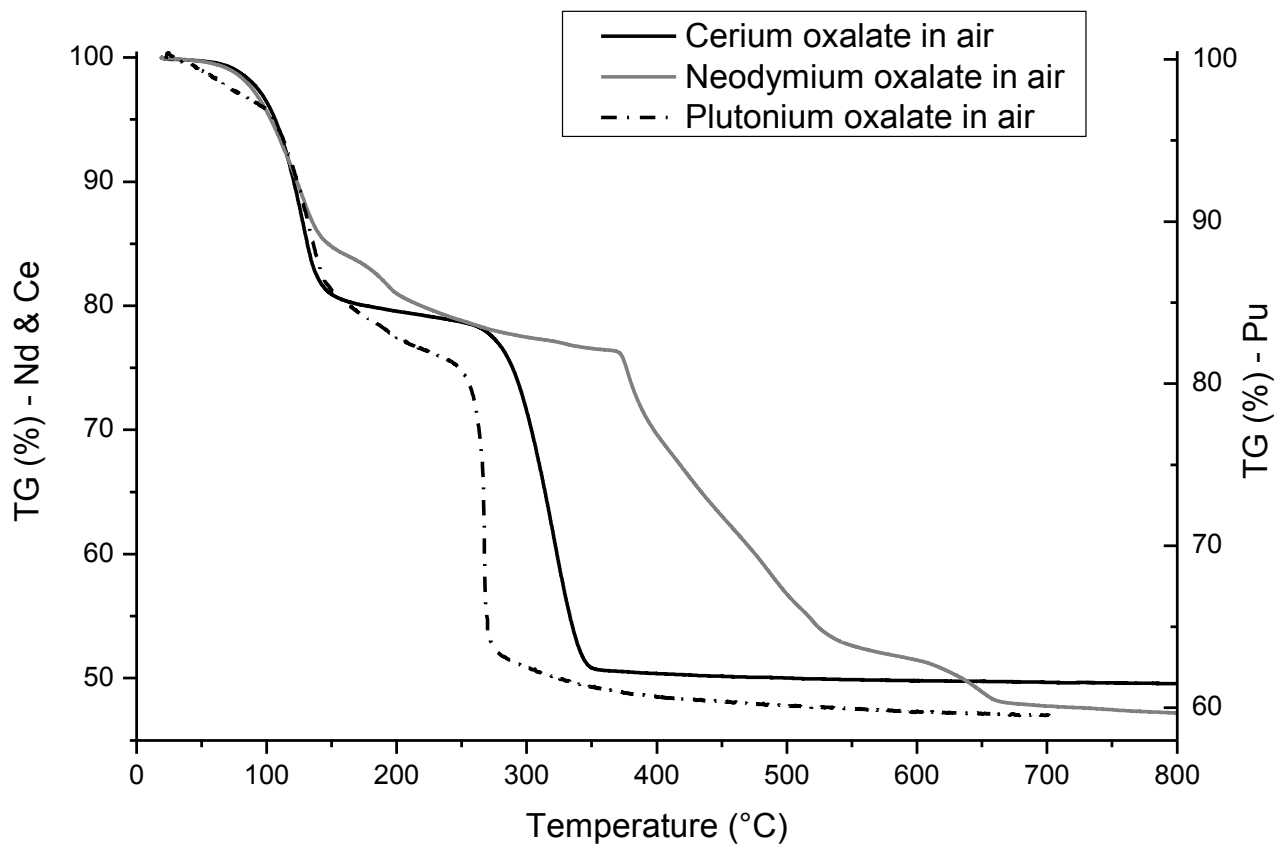

Figure 15 : Comparison between the thermal decomposition of neodymium, cerium and plutonium oxalates in air on heating at 2°C/min.

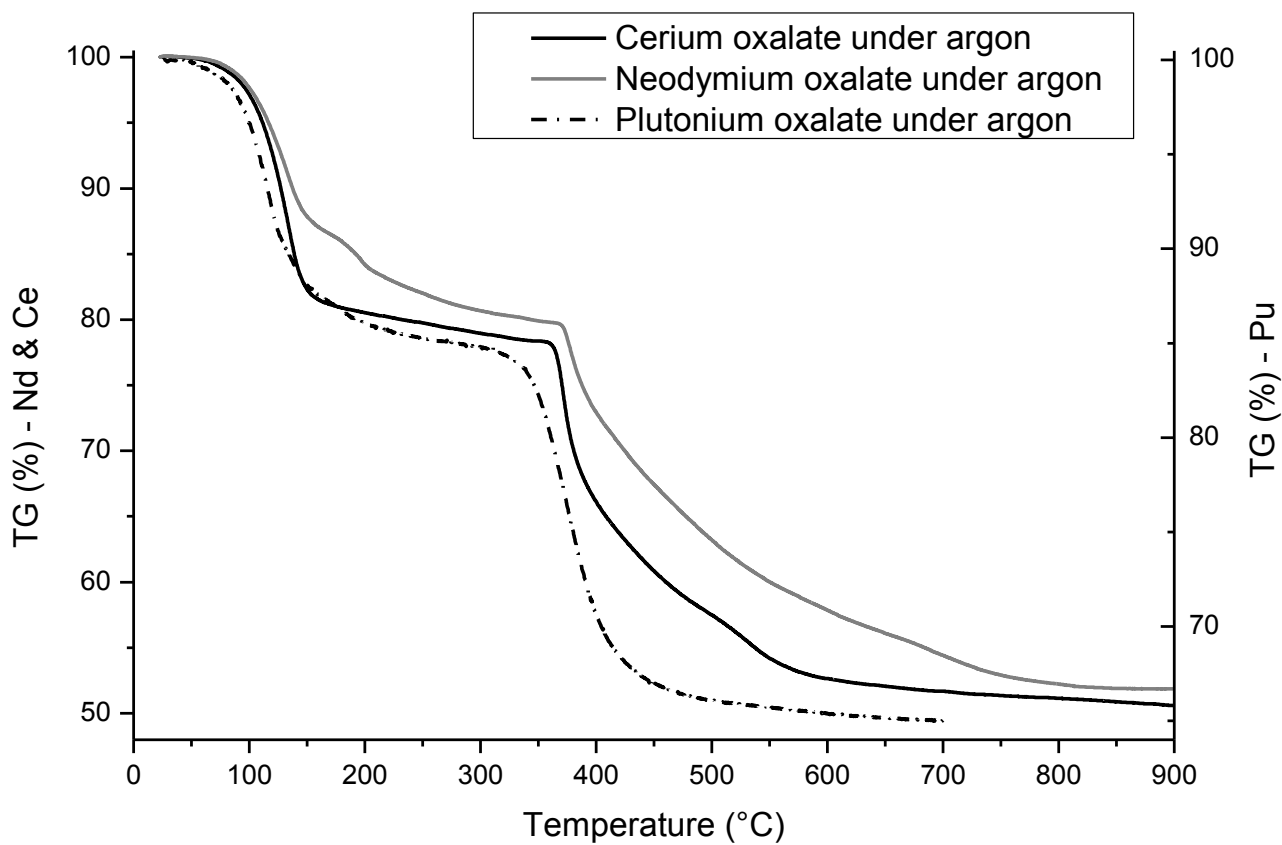

Figure 16 : Comparison between the thermal decomposition of neodymium, cerium and plutonium oxalates under argon on heating at 2°C/min.

The superimpositions of the TG curves corresponding to the calcination of neodymium, cerium and plutonium oxalate in air and under argon are shown on Figures 15 and 16 respectively, clearly pointing out the prevalence of the redox properties of the metal cation on the decomposition mechanism. In air, the thermal decomposition paths of cerium and plutonium oxalates are quite similar: after the dehydration phase, the oxalates decompose in one step, accelerated by the oxidation of the metal cation. Plutonium oxide is formed at lower temperature than cerium oxide, demonstrating that $Pu^{3+}$ is more easily oxidised than $Ce^{3+}$. Under argon, trivalent cerium is stabilised at higher temperature than trivalent plutonium, allowing the apparition of a trivalent carbonate intermediate for cerium but not for plutonium.

On the other hand, with these isostructural oxalate compounds, redox properties of the metal don't have a noticeable impact on the residual elemental carbon content in the oxides. Indeed, the carbon over metal ratio was approximately the same in neodymium and cerium oxides when the oxalate compounds were calcined under inert atmosphere. This result is quite unexpected and subsequent work is needed on this matter using other isostructural lanthanides or actinides oxalates.

## Conclusions

Thermal analyses and complementary techniques allowed us to characterise the thermal decomposition of two lanthanide and one actinide oxalates – namely neodymium, cerium and plutonium, all in the trivalent oxidation state.

The thermal behaviour of these compounds depends on the nature of the atmosphere imposed during the calcination (inert or oxidizing). While dehydration is always endothermic, decomposition of the oxalate groups appears either endothermic under argon or exothermic under air (only partially for neodymium oxalate). Decomposition per se is endothermic but other reactions superimpose (like the oxidation of carbon monoxide) resulting in an exothermic heat flow.

The thermal decomposition mechanisms were determined under air and argon for each compound, based on consecutive mass losses recorded by TG, gases emission profiles collected by MS or FT-IR, and characterisations of selected intermediates by HT-XRD and FT-IR. Heats of reactions corresponding to each elementary step were measured by DTA or DSC. Carbon content in the oxides obtained after calcination was assessed. Kinetic parameters for the lanthanide oxalates decomposition were calculated based on the TG results at three different heating rates, and appeared consistent with kinetic parameters previously published in the literature.

It emerged that some similarities can be noticed between the behaviour of plutonium and neodymium or cerium compounds:

- first, the dehydration of plutonium(III) oxalate occurs in several steps and leads to the formation of the dihydrate intermediate, like neodymium oxalate;
- the anhydrous oxalate is formed regardless of the metal cation nature;
- then, the decomposition of the oxalate groups occurs in one single step for plutonium(III) oxalate, as cerium oxalate under air;
- the enthalpies of dehydration are close for the three compounds (around 300-400 kJ/mol), as well as the decomposition enthalpies for cerium and plutonium(III) oxalates (around + 80 kJ/mol under argon or – 450-550 kJ/mol under air);
- like cerium, plutonium undergoes oxidation during the last step of the thermal decomposition, as the trivalent metal oxalate becomes the tetravalent metal oxide.

However, even if the compounds have the same stoichiometry, crystal structure, and the metal cations have the same oxidation state in the beginning (+III), they display distinctive singularities in their thermal decomposition path:

- regardless of the atmosphere plutonium(III) oxalate decompose into oxide in one step, while neodymium and cerium oxalate decomposition path depends strongly whether the atmosphere is oxidizing or inert: cerium oxalate goes through a carbonate intermediate under argon only, and neodymium oxalate forms one oxalato-carbonate intermediate in air plus two carbonate intermediates under both atmosphere;
- generally speaking, the conversion into oxide is easier for plutonium oxalate, regardless of the atmosphere, than for neodymium and cerium oxalates: the conversion to $PuO_2$ is achieved at 400°C under air and 750°C under argon, whereas the conversion to $CeO_2$ and $Nd_2O_3$ is achieved at 450°C and 700°C respectively under air and around 800-850°C for both lanthanides under argon.

The comparative examination of the thermal decomposition of neodymium, cerium and plutonium(III) oxalates revealed that their behaviour is noticeably distinct, each compound displaying special features. Nevertheless, a close connection can be made between the calcination of cerium and plutonium(III) oxalate under air, even if the dehydration phase is slightly different, and cerium oxalate can be considered as an acceptable model for plutonium(III) oxalate in this case.

Thus, using lanthanides as surrogates for actinide compounds should be carefully undertaken, keeping in mind that even if the starting compounds are similar, the nature of the metal cation prevails in determining the decomposition route.

## Experimental Section

**Preparation of the $Ln^{III}$ and $An^{III}$ oxalates:** The lanthanide compounds, $Nd_2(C_2O_4)_3 \cdot nH_2O$ and$(C_2O_4)_3 \cdot nH_2O$, were synthesised by mixing an acid $Ln^{III}$ nitrate solution (from cerium(III) nitrate hexahydrate, ALDRICH, 99% and neodymium(III) nitrate hexahydrate, Alfa Aesar, 99.9%) and a concentrated $H_2C_2O_4$ solution (from oxalic acid dihydrate, Prolabo, 99%) in a vortex effect reactor with a slight excess of $H_2C_2O_4$. The resulting crystallised powder were filtered off, washed with a 90/10 ethanol/water solution and dried at room temperature. The crystallographic structure of the compounds was confirmed by XRD (isomorphic, JCPDS 00-025-0567), and the compounds stoichiometry was verified by ICP-AES and acid-base titration.
For the $Pu_2(C_2O_4)_3 \cdot nH_2O$, a $Pu^{III}$ solution was prepared in a glove box by reducing a nitric solution of $Pu^{IV}$ (96.9% of $^{239}Pu$) with concentrated hydrazinium nitrate prepared by neutralizing carefully hydrazine (Sigma-Aldrich, 37w% in water) with nitric acid. It was then precipitated with a concentrated $H_2C_2O_4$ solution in a vortex reactor. The resulting precipitate was then filtered off, washed with a 90/10 ethanol/water solution and dried at room temperature after dividing the powder.
**Analytical techniques for lanthanide oxalates:** Thermogravimetric (TG) and differential thermal analyses (DTA) were carried out with a SETARAM Setsys 16/18 Supersonic analyzer under argon or air flow (250 mL/min), at a constant rate of 2, 5 and 10°C/min up to 1000°C. This instrument was coupled with two evolved gas analyzers: a Pfeiffer QMA 400 Mass Spectrometer and a Thermo Scientific Nicolet 380 FT-IR. Samples of oxalate weighing approximately 20-30 mg were placed in a 100 µL alumina crucible. Using the same procedure, solid intermediates were isolated at specific temperatures for immediate structural investigations.

X-ray powder diffraction (XRD) data at room temperature were acquired on a D8 BRUKER AXS diffractometer, and high-temperature diffraction studies were carried out with an X'Pert PRO MPD Panalytical diffractometer equipped with a temperature chamber under controlled atmosphere of air or nitrogen. Based on the scans in the range of $5° \leq \theta \leq 80°$, the d-spacing and relative intensities were obtained and matched with JCPDS database for phase identification purposes.

UV-vis spectra of the solid intermediates and products were acquired between 400 and 800 nm with a HITACHI U-3000 analyzer equipped with an integration sphere.

Infrared spectra (FT-IR) of all solid samples were recorded with a BRUKER Equinox 55 at a resolution of 4 cm$^{-1}$, over the spectral range of 400 to 4000 cm$^{-1}$.

SEM observations were performed using an environmental scanning electron microscope Quanta 600F BRUKER AXS equipped with an EDS system.

The carbon weight percentage within the compounds at different stages of their thermal decomposition was measured with a carbon-sulphur HORIBA EMIA-320V analyser.

**Analysis techniques for plutonium oxalates:** Thermogravimetric and differential scanning calorimetry (DSC) analyses were carried out with a NETZSCH STA 409C analyzer under argon flow (3 NL/h) or static air, at a constant rate of 2°C/min up to 700°C. Samples of oxalate weighing approximately 40-50 mg were placed in an 85 µL alumina crucible.

X-ray powder diffraction data were acquired with an INEL CPS 120 diffractometer, with silicon as an internal standard. Each solid sample was mixed into an epoxy resin in order to avoid the dispersion of the contamination.

UV-vis spectra of the solid products were acquired with the same apparatus as for the lanthanide oxalates. UV-vis spectra of the plutonium solutions were acquired with a double beam CINTRA 10e GBC spectrophotometer, with a working range of 300-1100 nm.

Infrared spectra of KBr discs containing the samples were recorded with a NICOLET MAGNA IR 550 series II. A spectral range from 400 to 4000 cm$^{-1}$ was employed.

**Acknowledgments**

The authors are thankful to J. Dauby, CEA, for helping with lanthanide oxalate syntheses, to O. Devisme, IJL, for helping with thermal analyses and SEM observations, and to G. Medjahdi and P. Villeger, IJL, for providing us with the high temperature X-ray diffraction data of the compounds.